\documentclass[12pt]{article}
\usepackage{graphicx}
\usepackage{amssymb}
\usepackage{amsmath}

\newcommand{\definition}[1]{

\setbox0=\hbox{\vbox to 20pt {}}

\setbox1=\hbox{\vbox{\parbox[l]{4.1in}{  \noindent \phantom{$\int\limits_0^1$} \hspace{-16pt}\vspace{-3pt} #1}}}

\setbox2=\hbox{\vbox{\hbox{\underline{Definition}: \phantom{$\int\limits_0^1$} \hspace{-24pt}}}}

\setbox3=\vbox to \ht1 {\box2 \vfill} 

\hbox{\box3 \box1}

\vspace{7pt}
}

\newcommand{\tang}[1]{\underline{#1}}

\newcommand{\strike}[1]{
\setbox0=\hbox{$#1$}
\skip0 = \wd0
\setbox1=\vbox{}
\advance\skip0 by 1pt
\skip1 = \ht1
\advance\skip1 by 3pt
#1  \hskip -\skip0 \diagup
}

\begin{document}

\begin{titlepage}
\begin{flushright}
CALT-68 2722
\end{flushright}

\begin{center}
{\Large\bf $ $ \\ $ $ \\
Symmetries of massless vertex operators in $AdS_5\times S^5$
}\\
\bigskip\bigskip\bigskip
{\large Andrei Mikhailov}
\\
\bigskip\bigskip
{\it California Institute of Technology 452-48,
Pasadena CA 91125 \\
\bigskip
and\\
\bigskip
Institute for Theoretical and 
Experimental Physics, \\
117259, Bol. Cheremushkinskaya, 25, 
Moscow, Russia}\\

\vskip 1cm
\end{center}

\begin{abstract}
The worldsheet sigma-model of the superstring in $AdS_5\times S^5$ has a
one-parameter family of flat connections parametrized by the spectral
parameter. The corresponding Wilson line is not BRST invariant for an open
contour, because the BRST transformation leads to boundary terms. These
boundary terms define a cohomological complex associated to the endpoint of the
contour.  We study the cohomology of this complex for Wilson lines in some
infinite-dimensional representations.  We find that for these representations
the cohomology is nontrivial at the ghost number 2. This implies that it is
possible to define a BRST invariant open Wilson line. 
The central point in the construction is the existence of massless 
vertex operators 
transforming exactly covariantly under the action of the global symmetry 
group. In flat space massless vertices transform covariantly only up to adding
BRST exact terms. But in AdS we show that it is possible to define vertices 
so that they transform exactly covariantly.
\end{abstract}

\end{titlepage}

\section{Introduction}
Nonlocal conserved charges play the central role in quantum integrability \cite{Luscher:1977uq}.
For the superstring in $AdS_5\times S^5$ their existence was proven in
the classical sigma-model in \cite{Bena:2003wd} using the  Green-Schwarz-Metsaev-Tseytlin
formalism, and in \cite{Vallilo:2003nx,Berkovits:2004jw,Berkovits:2004xu} using
the pure  spinor formalism. The existence of the nonlocal conserved charges
at the quantum level was proven in \cite{Berkovits:2004xu,Puletti:2008ym}.

\subsection{Open Wilson lines on the worldsheet}
\subsubsection{Wilson lines and BRST operator}
The most important feature of the string  
worldsheet $\sigma$-model (besides the conformal 
invariance) is the existence of the BRST structure. 
The physically meaningful constructions
should respect the action of the BRST operator $Q_{BRST}$. 

The nonlocal conserved charges of \cite{Vallilo:2003nx,Berkovits:2004jw,Berkovits:2004xu}
are only BRST invariant up to the boundary terms. 
To be more precise, we need
to introduce the transfer matrix which is the generating function of the nonlocal
conserved charges.
For an open contour $C_A^B$ connecting points $A$ and $B$ on the string worldsheet:

\vspace{10pt}

\hspace{0.8in}\includegraphics[width=2.4in]{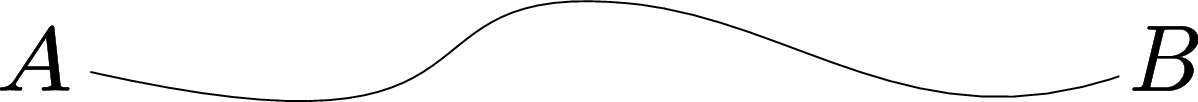}

\vspace{10pt}

\noindent
we define the transfer matrix:
\begin{equation}\label{TransferMatrixDefinition}
T_{\rho}[C_A^B] = P\exp \left( -\int^B_A J[z] \right)
\end{equation}
where $C_A^B$ is a contour connecting points $A$ and $B$ on the string worldsheet,
and $\rho$ is a representation of ${\bf g}=psu(2,2|4)$. The currents $J[z]$ depend
on the spectral parameter $z$ and the representation $\rho$. 
The transfer matrix is a function of the {\em spectral parameter} $z$. It is
the generating function of the conserved charges.
It can be also though of as the Wilson line operator corresponding to 
the flat connection $J[z]$ on the string worldsheet. In this paper we will use 
both expressions: ``transfer
matrix'' and ``Wilson line'', and understand them as synonyms.

The BRST variation of the Wilson line $T_{\rho}[C_A^B]$ results in the boundary terms. 
Using the  notations
of \cite{Mikhailov:2007eg}:
\begin{eqnarray}\label{BoundaryTermsInBRST}
\varepsilon Q_{BRST} T[C_A^B](z) & = &
\left( {1\over z} \varepsilon\lambda_3(B) + z \varepsilon\lambda_1(B) \right) 
T[C_A^B](z) -
\nonumber
\\
&& - T[C_A^B](z) \left( {1\over z} \varepsilon\lambda_3(A) + z \varepsilon\lambda_1(A) \right) 
\end{eqnarray}
This equation is very interesting. Nontrivial boundary terms in (\ref{BoundaryTermsInBRST}) allow
to ``bootstrap'' at least at the classical level the  structure of $r$---$s$ matrices,
see Section  7 of \cite{Mikhailov:2007eg}.
On the other hand, these boundary terms present a problem: 
the nonlocal conserved charges are not
physical quantities, at least not  in an obvious sense. 
(Because they are not in the kernel of $Q_{BRST}$.) What should we do with them?

\subsubsection{The plugs}
\label{sec:ThePlugs}
A natural thing to try is to find an operator which when inserted at the endpoint
of the
Wilson line would make it $Q_{BRST}$-closed. 
This is somewhat analogous to the open Wilson line in QCD. The expression
$P\exp\int_A^B A_{\mu}dx^{\mu}$ is unphysical, because it is not gauge invariant.
The physical quantity is:
\begin{equation}
\overline{\psi}(B) \left( P\exp\int_A^B A_{\mu}dx^{\mu} \right) \psi(A)
\end{equation}
We may call $\psi(A)$ and $\overline{\psi}(B)$ ``the plugs'' because they fix
``leaking boundary terms'' in gauge tranformations, or in BRST transformations.
Can we find similar plugs  for the Wilson
line  on the string worldsheet in $AdS_5\times S^5$?
In this  paper we will report a progress in this direction. 

\subsubsection{Wilson lines in infinite-dimensional representations}
\label{sec:WilsonLinesInInfDimRep}
 Remember that the
Wilson line depends on a choice of representation; we have to choose a
representation $\rho$ of $psu(2,2|4)$.  Consider the space of states of the
linearized supergravity multiplet in $AdS_5\times S^5$. It splits into the
direct sum of infinitely many infinite-dimensional irreducible representations
of $psu(2,2|4)$, each corresponding to a BPS state.  We will argue that when
$\rho$ is one of those infinite-dimensional BPS representations, then there is
a suitable plug of the ghost number 2. This is closely related to the vertex
operators for the massless states in $AdS_5\times S^5$.  In fact we will
relate the BRST cohomology complex corresponding to the endpoint of the Wilson
line to the BRST complex corresponding to the vertex operators. This is
essentially an example of the  Frobenius reciprocity. The main
nontrivial point is the construction of the vertex operators transforming {\em
strictly covariantly} under the global supersymmetries of $AdS_5\times S^5$.
This is different from flat space where massless vertices transform covariantly
only  up to BRST exact terms. 

Wilson lines in infinite-dimensional representations played an important
role in the integrable context in \cite{Bazhanov:1996dr} in the construction
of the Q-operator. They also played an important role in the AdS/CFT context
in \cite{Berkovits:2008qc} for the interpretation of the YM Feynman diagramms
in the string worldsheet theory.

\subsection{Representation theory interpretation of the SUGRA spectrum}
Consider an infinite-dimensional irreducible representation 
${\cal H}$ of $psu(2,2|4)$. It is natural to ask the question: 
\begin{align}
& \mbox{When does ${\cal H}$ appear in the decomposition of the space
of  solutions} 
\nonumber \\ 
& \mbox{of the linearlized Type IIB SUGRA equations 
in $AdS_5\times S^5$?}
\end{align}
The results of our paper imply that this answer can be answered
directly in terms of the structure of ${\cal H}$, as a representation
of $psu(2,2|4)$. Namely, given ${\cal H}$, we consider the following complex:
\begin{equation}
\ldots \longrightarrow 
{\cal H}'\otimes_{{\bf g}_{\bar{0}}} {\cal P}^n 
\stackrel{Q_{endpoint}}{\longrightarrow}
{\cal H}'\otimes_{{\bf g}_{\bar{0}}} {\cal P}^{n+1} 
\longrightarrow \ldots
\end{equation}
defined entirely in terms of ${\cal H}$ --- see Section 
\ref{sec:DefinitionOfAlgebraicComplex}. Then, we claim that the multiplicity of ${\cal H}$ in the space of
linearized SUGRA solutions is equal to the
dimension of the second cohomology 
$H^2(Q_{endpoint}\;,\; {\cal H}'\otimes_{{\bf g}_{\bar{0}}}{\cal P}^{\bullet})$ of this
complex. (We only claim this when ${\cal H}$ has ``high enough''
spin on $S^5$; see the end of Section \ref{sec:CovariantUniversal}.)

\subsection{The plan of the paper}
Most of the paper is about massless vertex operators in $AdS_5\times S^5$.
In Section \ref{sec:VerticesInGroupTheoryLanguage} we give a geometrical
definition of vertex operators using the representation of $AdS_5\times S^5$
as the coset space $G/H$. 
In Section \ref{sec:DefinitionOfCovariantVertex} we explain what it means
for the vertex to be strictly covariant, and then in Sections 
\ref{sec:TaylorSeries} and \ref{sec:CohomologicalArgument}
prove the existence of such covariant vertices. 
In Section \ref{sec:FlatSpaceLimit} we discuss the flat space limit of our
construction. 
(Although it is impossible to  construct the strictly covariant vertex
in the flat space, but nevertheless the construction in $AdS_5\times S^5$
has a well-defined flat space limit, which does transforms strictly  
covariantly, but only under a subgroup  $SO(1,4)\times SO(5)\subset SO(1,9)$.)
In Section \ref{sec:RelationToEndpoint} we explain how the covariant vertex
plugs the endpoint of the Wilson line.
In Section \ref{sec:Applications} we present some consequences of our construction;
we explain how to prepare the vertex operator depending on the spectral parameter.

\subsection{Notations}
\label{sec:Notations}
The algebra of supersymmetries of $AdS_5\times S^5$ has a ${\bf Z}_4$ grading:
\begin{equation}
psu(2,2|4) = {\bf g} = {\bf g}_{\bar{0}} + {\bf g}_{\bar{1}} + {\bf g}_{\bar{2}}
+{\bf g}_{\bar{3}}
\end{equation}
We denote ${\cal U}{\bf g}$ the universal enveloping algebra of ${\bf g}$.

\vspace{5pt}

\noindent Let $g$ denote the group element of ${\cal U}{\bf g}$, {\it i.e.}
an element of the group $PSU(2,2|4)$:
\begin{equation}
g=e^{\omega} e^{\theta_L+\theta_R} e^x
\end{equation}
where $\omega\in {\bf g}_{\bar{0}}$, $\theta_L\in {\bf g}_{\bar{3}}$,
$\theta_R\in {\bf g}_{\bar{1}}$ and $x\in {\bf g}_{\bar{2}}$.
So defined $\theta_{L,R}$ and $x$ are coordinates of the
super-$AdS_5\times S^5$.
The generators of ${\bf g}=psu(2,2|4)$ are the same as
in \cite{Mikhailov:2007mr,Mikhailov:2007eg}:
\begin{equation}
t^3_{\alpha} \in {\bf g}_{\bar{3}},\;\;
t^1_{\dot{\alpha}} \in {\bf g}_{\bar{1}},\;\;
t^2_{\mu} \in {\bf g}_{\bar{2}},\;\;
t^0_{[\mu\nu]} \in {\bf g}_{\bar{0}}
\end{equation}

\vspace{5pt}

\noindent
For a vector space $L$ we will denote $L'$ the dual vector space. In particular,
the space of states is denoted ${\cal H}$ and the space of linear functionals
on the states is denoted ${\cal H}'$. We will mostly consider the
linear functionals which are the values of some supergravity fields
(such as the Ramond-Ramond field strength) at a fixed point in $AdS_5\times S^5$.
These could be also thought of as non-normalizable elements of ${\cal H}$, 
``delta-functions type of states'' in ${\cal H}$.

\vspace{5pt}
\noindent
For an even vector space $L$ we denote $\Lambda^n L$ the space of antisymmetric
tensors. For an odd vector space $\Lambda^n L$ will stand for symmetric tensors.

\vspace{5pt}
\noindent Example: ${\bf g}_{\bar{3}}$ is an odd vector space, and ${\bf
g}'_{\bar{3}}$  is the dual space. Therefore $\Lambda^5 {\bf g}'_{\bar{3}}$ is
identified with the fifth order polynomials of some bosonic spinor variable
$\lambda^{\alpha}$.

\section{Massless vertex operators as functions on the group manifold}
\label{sec:VerticesInGroupTheoryLanguage}

Massless vertex operators in $AdS_5\times S^5$ were introduced
in \cite{Berkovits:2000yr}.

Using the group theory language, we can define the vertex operator as a collection
of functions $V_{\alpha\beta}(g)$, $V_{\alpha\dot{\beta}}(g)$ 
and $V_{\dot{\alpha}\dot{\beta}}(g)$ of $g\in PSU(2,2|4)$ subject to the condition of
${\bf g}_0$-covariance, which says that for any $h\in SO(1,4)\times SO(5)$ we should get:
\begin{eqnarray}
V_{\alpha\beta}(hg) & = & h_{\alpha}^{\alpha'} h_{\beta}^{\beta'} V_{\alpha'\beta'}(g)
\label{VertexCovariance1}
\\
V_{\alpha\dot{\beta}}(hg) & = & h_{\alpha}^{\alpha'} h_{\dot{\beta}}^{\dot{\beta}'}
V_{\alpha'\dot{\beta}'}(g)
\label{VertexCovariance2}
\\
V_{\dot{\alpha}\dot{\beta}}(hg) & = & 
h_{\dot{\alpha}}^{\dot{\alpha}'} h_{\dot{\beta}}^{\dot{\beta}'}
V_{\dot{\alpha}'\dot{\beta}'}(g)
\label{VertexCovariance3}
\end{eqnarray}
Here $h_{\alpha}^{\alpha'}$ and $h_{\dot{\beta}}^{\dot{\beta}'}$ 
are the matrix elements of $h\in G_0$ acting on ${\bf g}_3$ and ${\bf g}_1$
respectively.

For a tensor field $\varphi_{\alpha_1\ldots\alpha_m\;\dot{\beta}_1\ldots\dot{\beta}_n}(g)$ 
we introduce the covariant derivatives:
\begin{eqnarray}
{\cal T}^{\bar{3}}_{\alpha} \varphi_{\alpha_1\ldots\alpha_m\;\dot{\beta}_1\ldots\dot{\beta}_n}(g) 
& = & \left.{d\over ds}\right|_{s=0} 
\varphi_{\alpha_1\ldots\alpha_m\;\dot{\beta}_1\ldots\dot{\beta}_n}(e^{-s t^3_{\alpha}}g)
\nonumber
\\
{\cal T}^{\bar{2}}_m \varphi_{\alpha_1\ldots\alpha_m\;\dot{\beta}_1\ldots\dot{\beta}_n}(g) & = & 
\left.{d\over ds}\right|_{s=0} 
\varphi_{\alpha_1\ldots\alpha_m\;\dot{\beta}_1\ldots\dot{\beta}_n}(e^{-s t^2_m}g)
\label{DefCovariantDerivatives}
\\
{\cal T}^{\bar{1}}_{\dot{\alpha}} \varphi_{\alpha_1\ldots\alpha_m\;\dot{\beta}_1\ldots\dot{\beta}_n}(g) & = & 
\left.{d\over ds}\right|_{s=0} 
\varphi_{\alpha_1\ldots\alpha_m\;\dot{\beta}_1\ldots\dot{\beta}_n}(e^{-s t^1_{\dot{\alpha}}}g)
\nonumber
\\
{\cal T}^{\bar{0}}_{[mn]} \varphi_{\alpha_1\ldots\alpha_m\;\dot{\beta}_1\ldots\dot{\beta}_n}(g) & = & 
\left.{d\over ds}\right|_{s=0} 
\varphi_{\alpha_1\ldots\alpha_m\;\dot{\beta}_1\ldots\dot{\beta}_n}(e^{-s t^0_{[mn]}}g)
\nonumber
\end{eqnarray}
Collectively:
\begin{equation}\label{Collectively}
{\cal T}^{\overline{n}}_A \varphi_{\alpha_1\ldots\alpha_m\;\dot{\beta}_1\ldots\dot{\beta}_n}(g)  =  
\left.{d\over ds}\right|_{s=0} 
\varphi_{\alpha_1\ldots\alpha_m\;\dot{\beta}_1\ldots\dot{\beta}_n}(e^{-s t^{\overline{n}}_A}g)
\end{equation}
Note that the covariant derivatives ${\cal T}_A^{\bar{n}}$ for $\bar{n}\neq \bar{0}$ satisfy the 
condition of ${\bf g}_0$-covariance, for example:
\begin{equation}\label{CovarianceOfCovariantDerivatives}
{\cal T}^3_{\alpha} V_{\beta\gamma}(hg)  =
h_{\alpha}^{\alpha'}
h_{\beta}^{\beta'}
h_{\gamma}^{\gamma'} 
{\cal T}^3_{\alpha'} V_{\beta'\gamma'}(g)
\end{equation}
Also, the ${\bf g}_0$-covariance condition 
(\ref{VertexCovariance1}) --- (\ref{VertexCovariance3}) can be
formulated as the following explicit expression for the covariant derivative
${\cal T}_{[\mu\nu]}$ along ${\bf g}_0$:
\begin{equation}
{\cal T}^0_{[\mu\nu]} V_{\beta\gamma}(g) = 
\left.{d\over ds}\right|_{s=0} V_{\beta\gamma}(e^{-s t^0_{[\mu\nu]}} g)
= {f_{[\mu\nu]\beta}}^{\beta'} V_{\beta'\gamma}(g) + 
  {f_{[\mu\nu]\gamma}}^{\gamma'} V_{\beta\gamma'}(g)
\end{equation}
The condition that the vertex operator is ${\cal Q}$-closed can be written as follows:
\begin{eqnarray}
{\cal T}^3_{(\alpha} V_{\beta\gamma)} & = & {f_{(\alpha\beta}}^m S_{\gamma)m}
\label{QVZeroA}
\\
{\cal T}^1_{(\dot{\alpha}} V_{\dot{\beta}\dot{\gamma})} & = &
{f_{(\dot{\alpha}\dot{\beta}}}^m S_{\dot{\gamma})m}
\label{QVZeroB}
\\
{\cal T}^3_{(\alpha} V_{\beta)\dot{\gamma}} +
{\cal T}^1_{\dot{\gamma}} V_{\alpha\beta} & = & {f_{\alpha\beta}}^m A_{m\dot{\gamma}}
\label{QVZeroC}
\\
{\cal T}^1_{(\dot{\alpha}|} V_{\gamma | \dot{\beta})} +
{\cal T}^3_{\gamma} V_{\dot{\alpha}\dot{\beta}} 
& = & 
{f_{\dot{\alpha}\dot{\beta}}}^m A_{\gamma m}
\label{QVZeroD}
\end{eqnarray}
where $A$ and $S$ are defined by these equations.

The gauge transformations are:
\begin{eqnarray}
\delta_{\Phi,\widetilde{\Phi}} V_{\alpha\beta} & = &  {\cal T}^3_{(\alpha} \Phi_{\beta)}
\\
\delta_{\Phi,\widetilde{\Phi}} V_{\dot{\alpha}\dot{\beta}} & = & 
{\cal T}^1_{(\dot{\alpha}} \widetilde{\Phi}_{\dot{\beta})}
\\
\delta_{\Phi,\widetilde{\Phi}} V_{\alpha\dot{\beta}} & = & 
{\cal T}^3_{\alpha} \widetilde{\Phi}_{\dot{\beta}} +
{\cal T}^1_{\dot{\beta}} \Phi_{\alpha} 
\end{eqnarray}
where $\Phi$ and $\widetilde{\Phi}$ are the parameters of the gauge transformations.

The BRST operator is:
\begin{equation}\label{LocalComplex}
{\cal Q} = {\cal Q}_L + {\cal Q}_R = \lambda^{\alpha}{\cal T}^3_{\alpha} 
+\tilde{\lambda}^{\dot{\alpha}} {\cal T}^1_{\dot{\alpha}}
\end{equation}
Our definition of the vertex is slightly weaker than the definition
of \cite{Berkovits:2000yr}. The definition of \cite{Berkovits:2000yr}
requires that: 
\begin{equation}\label{GhostNumberGauge}
V_{\alpha\beta}=0 \;\;\mbox{and }\; V_{\dot{\alpha}\dot{\beta}}=0
\end{equation}
the only nonzero component of the vertex remains 
$V_{\alpha\dot{\beta}}$. In fact $V_{\alpha\beta}$ and
$V_{\dot{\alpha}\dot{\beta}}$ are always $Q$-exact. Therefore
the condition (\ref{GhostNumberGauge}) can always be satisfied by
adding to the vertex something ${\cal Q}_{BRST}$-exact. 
In this sense the components $V_{\alpha\beta}$ and 
$V_{\dot{\alpha}\dot{\beta}}$ can always be ``gauged away''.
But we want the {\em covariant} vertex operator. 
We suspect that it might be impossible to gauge away
$V_{\alpha\beta}$ and  $V_{\dot{\alpha}\dot{\beta}}$ in a covariant way.
This is the reason why we prefer to allow these components in
the definition of the vertex operator.

\section{Covariant vertex: the definition}
\label{sec:DefinitionOfCovariantVertex}

\subsection{Vertices and states}
In string theory  vertex operators represent states. 
Vertex operators are functions $V(x,\theta,\lambda)$.
The global symmetries act
on $x$ and $\theta$, and therefore act on vertex operators.

On the other hand, the global symmetries act on the space of states.
Therefore the action of the global symmetry group on states
should agree with the action on vertex operators. 
Naively, this would imply that if $V_{\Psi}(x,\theta,\lambda)$ is
a vertex operator corresponding to the state $\Psi$ then
\begin{equation}\label{TransformsCovariantly}
V_{g\Psi}(x,\theta,\lambda)=V_{\Psi}(gx,g\theta,\lambda)
\end{equation}
But in fact this formula, generally speaking, holds only up to
BRST-trivial corrections (terms which are $Q_{BRST}$ of something).

Note that $V_{\Psi}(x,\theta,\lambda)$ is not defined unambiguously,
because we could add to it BRST-exact terms and get physically
equivalent vertex.
We will prove that in $AdS_5\times S^5$ it is possible
to use this freedom in the definition of $V_{\Psi}$ and choose 
$V_{\Psi}$ so that it transforms covariantly, as in (\ref{TransformsCovariantly}).
In our proof we will use the fact that vertices corresponding to
supergravity states {\em exist}. This was proven in \cite{Berkovits:2000yr}.
Given the existence of the vertex, we will prove that it is always possible
to correct it by a BRST-exact expression, if necessary, to get a
{\em covariant} vertex.

\subsection{Examples of covariant vertices: AdS radius and $\beta$-deformation}
The first example of the covariant vertex was given in 
\cite{Berkovits:2008ga}. It was shown that the {\em zero mode} dilaton
vertex is given by the expression\footnote{Using the notations of
\cite{Berkovits:2008ga}: 
$\eta_{\alpha\hat{\alpha}}\lambda^{\alpha}\hat{\lambda}^{\hat{\alpha}}$}
which is independent of $x$ and $\theta$:
\begin{equation}\label{ZeroModeVertex}
V(x,\theta,\lambda) = \mbox{Str}(\lambda_3\lambda_1)
\end{equation}
This operator plays the central role in \cite{Berkovits:2008ga}.
The corresponding marginal deformation of the action changes the
radius of $AdS_5\times S^5$. The radius is invariant under the
global symmetries, therefore in this case covariance means invariance;
the vertex (\ref{ZeroModeVertex}) is invariant under the global symmetries.
This is related to the fact that the action of the pure spinor superstring
in $AdS_5\times S^5$ is exactly invariant under the global symmetries
(while the action in flat space is invariant only up to adding a total
derivative).
Our construction can be considered a generalization of (\ref{ZeroModeVertex})
for fields with nontrivial dependence on $x$ and $\theta$.

The second example\footnote{Note in revised version: a more detailed
discussion of $V^{beta}_{ab}$ will be presented in the forthcoming
paper with O.~Bedoya, L.~Bevil\'{a}qua, 
and V.O.~Rivelles} is:
\begin{equation}\label{VBeta}
V^{beta}_{ab} = 
(g^{-1}\varepsilon(\lambda_3-\lambda_1)g)_a \;
(g^{-1}\varepsilon'(\lambda_3-\lambda_1)g)_b
\end{equation}
Here the indices $a$ and $b$ enumerate the adjoint representation
of $psu(2,2|4)$. Notice that (\ref{VBeta}) is antisymmetric under
the exchange of $a$ and $b$. Therefore this vertex transforms
in the antisymmetric product of two adjoint representations
of $psu(2,2|4)$. This antisymmetric product splits into two irreducible
components. The first component is the adjoint representation. But the 
part of (\ref{VBeta}) belonging to the adjoint representation is
actually  $Q_{BRST}$-exact:
\begin{eqnarray}
{f^{ab}}_c V^{beta}_{ab} =
[(g^{-1}\varepsilon(\lambda_3-\lambda_1)g),
(g^{-1}\varepsilon'(\lambda_3-\lambda_1)g)]_c =
\nonumber
\\
= \varepsilon Q_{BRST} (g^{-1} \varepsilon'(\lambda_3 + \lambda_1) g)_c
\end{eqnarray}
(Notice that this formula played an important role in 
Section 6 of \cite{Berkovits:2004xu}.)

The deformation of the action corresponding to (\ref{VBeta}) follows from the
standard descent procedure.  Let us denote: 
\begin{equation}
\Lambda_a(\varepsilon) = (g^{-1} \varepsilon( \lambda_3 - \lambda_1 ) g)_a
\end{equation}
This is the ghost number 1 cocycle corresponding to the
local conserved currents, see also Appendix \ref{sec:GhostNumberOne}.
It corresponds to the local conserved currents in the following
sense:
\begin{equation}
d\Lambda_a(\varepsilon) = \varepsilon Q (j_a)
\end{equation}
where $j_{a\pm}(\tau^+,\tau^-)$ is the density of the local
conserved charge corresponding to the global symmetries. Therefore:
\begin{equation}
d(\Lambda_{[a}(\varepsilon) \Lambda_{b]}(\varepsilon')) =
2\varepsilon Q j_{[a} \Lambda_{b]}(\varepsilon')
\end{equation}
and:
\begin{equation}
d(j_{[a}\Lambda_{b]}(\varepsilon)) =
-{1\over 2} \varepsilon Q (j_{[a}\wedge j_{b]})
\end{equation}
We conclude that for any constant antisymmetric matrix $B^{ab}$
we can infinitesimally deform the worldsheet action as follows:
\begin{equation}\label{LinearizedBeta}
S\to S + B^{ab} \int j_{[a}\wedge j_{b]}
\end{equation}
Consider for example $B_{ab}$ in the directions of $S^5$.
We get:
\begin{equation}\label{BInSDirections}
S\to S + B^{[kl][mn]} \left(
\int X_{[k} dX_{l]} \wedge X_{[m} dX_{n]}
+\ldots \right)
\end{equation}
where $X_j$ describes the embedding of $S^5$ into ${\bf R}^6$
and dots denote $\theta$-dependent terms. These $\theta$-dependent
terms appear because  $j_a$ includes $\theta$. 
Eq. (\ref{BInSDirections}) corresponds to the marginal
deformations of the ${\cal N}=4$ Yang-Mills known as 
$\beta${\em-deformations} \cite{Leigh:1995ep}, as follows from their quantum
numbers. 

The subspace
${\bf g}\subset {\bf g}\wedge {\bf g}$ corresponds to $B$ of the
following form:
\begin{equation}
B^{[kl][mn]} = \delta^{km} A^{ln} - \delta^{lm} A^{kn}
+ \delta^{ln} A^{km} - \delta^{kn} A^{lm}
\end{equation}
where $A^{mn}$ is antisymmetric matrix; then the corresponding
deformation of the Lagrangian is a total derivative
$d(A^{mn}X_m dX_n)$. The complementary space has real dimension
90, it corresponds to the representation $\bf 45_C$ of $so(6)$.
This is the expected quantum numbers of the linearized $\beta$-deformation, 
cp. Section 3.1
of \cite{Aharony:2002hx} and references therein. It was observed in
\cite{Aharony:2002hx} that some of these deformations are obstructed
when we pass from the linearized supergravity equations to the
nonlinear equations. Not all of the
deformations (\ref{LinearizedBeta}) can be extended to the solutions
of the nonlinear supergravity equations {\em as solutions constant
in $AdS_5$ directions}, but only those which satisfy 
some nonlinear equations on $B^{ab}$. If these nonlinear
equations are not satisfied, then the nonlinear solutions
will have ``resonant terms'' and because of these resonant
terms will not be periodic\footnote{Similar phenomenon for
``fast moving strings'' was discussed in 
\cite{Mikhailov:2004qf,Mikhailov:2005zd}.
Generally speaking, deviations
from periodicity in the global time of AdS correspond to something
like anomalous dimension. In this case it shows that the beta
function of the deformed theory is actually nonzero at the higher
order in the deformation, unless if additional (cubic) constraints are
imposed on the deformation parameter.}
in the global time of $AdS_5$.

\subsection{Universal vertex}
\label{sec:UniversalVertex}
Suppose that we are looking at the {\em massless} states transforming in some representation
${\cal H}$ of the global symmetry group $PSU(2,2|4)$.
For every state $\Psi\in {\cal H}$ we have the corresponding vertex operator
${\cal V}(\Psi)$. As we discussed, ${\cal V}(\Psi)$ consists of the components:
${\cal V}_{\alpha\beta}(\Psi)$, ${\cal V}_{\alpha\dot{\beta}}(\Psi)$ and
${\cal V}_{\dot{\alpha}\dot{\beta}}(\Psi)$. We write ${\cal V}$ instead of $V$
to stress that ${\cal V}$ is a function of $\Psi$. Mathematically 
it would be more appropriate to call it ``a linear operator from the space
of states to the space of vertex operators''. We will call ${\cal V}$ the
``universal vertex'' for the representation ${\cal H}$ because it is 
a uniform definition of vertex operators for all states in ${\cal H}$:
\[
{\cal V}\;:\;\; {\cal H} \longrightarrow \mbox{(functions of $x,\theta,\lambda$)}
\]
The global symmetry group $G=PSU(2,2|4)$ acts on both space of states and
space of vertex operators. It acts on the space of states by definition,
because it is the global symmetry group of the theory. It also acts on the
space of vertex operators. The action on the space of vertex operators
may seem obvious, but we would like to spell it out explicitly because we feel
that some confusion is possible. A vertex operator has components 
$V_{\alpha\beta}$, $V_{\alpha\dot{\beta}}$ and $V_{\dot{\alpha}\dot{\beta}}$
which are all functions of the group element $g$, {\it i.e.}
$V_{\alpha\beta}(g)$, $V_{\alpha\dot{\beta}}(g)$ and $V_{\dot{\alpha}\dot{\beta}}(g)$
satisfying the conditions of
${\bf g}_0$-covariance (\ref{VertexCovariance1}) --- (\ref{VertexCovariance3}).
Then, the action of the global symmetry transformation $g'\in PSU(2,2|4)$ is defined
as follows:
\begin{eqnarray}
(g'.V_{\alpha\beta})(g) & = & V_{\alpha\beta}(gg') \nonumber \\
(g'.V_{\alpha\dot{\beta}})(g) & = & V_{\alpha\dot{\beta}}(gg') 
\label{GlobalActionOnV} \\
(g'.V_{\dot{\alpha}\dot{\beta}})(g) & = & V_{\dot{\alpha}\dot{\beta}}(gg')
\nonumber
\end{eqnarray}
Because $g'$ hits $g$ on the right, this action of the global symmetries
is manifestly consistent with the conditions of
${\bf g}_0$-covariance (\ref{VertexCovariance1}) --- (\ref{VertexCovariance3})
and also commutes with the covariant derivatives  (\ref{DefCovariantDerivatives}).

This defines the action of $G$ on ${\cal V}(\Psi)$ for any fixed $\Psi$;
the expression: 
\begin{equation}
g'.({\cal V}(\Psi))
\label{GlobalActionOnUniversalVertex}
\end{equation}
is defined by (\ref{GlobalActionOnV}):
\begin{equation}
(\;\; g'.({\cal V}(\Psi))\;\; )(g) = (\;\; {\cal V}(\Psi) \;\; )(gg')
\end{equation}

\subsection{Covariant universal vertex}
It is natural to ask, if it is true that (\ref{GlobalActionOnUniversalVertex})
is equal to this:
\begin{equation}
{\cal V}(g'\Psi)
\end{equation}
In other words, if it is true or not that:
\begin{equation}\label{ExactCovariance}
(\; {\cal V}(\Psi) \; )(gg') \stackrel{?}{=} 
(\; {\cal V}(g'\Psi) \; )(g)
\end{equation}
This is not automatically true. What {\em is} automatically true\footnote{``Automatically true'' 
means true under the assumption that the vertex operators exist. The existence
was proven in \cite{Berkovits:2000yr}.} is this statement:
\begin{equation}\label{UpToBRSTExact}
(\; {\cal V}(\Psi) \; )(gg') = 
(\; {\cal V}(g'\Psi) \; )(g) \; + \; {\cal Q}_{BRST}(\mbox{smth})
\end{equation}
Remember that vertex operators are defined modulo
BRST-exact expressions. The question is, can we choose a representative for ${\cal V}(\Psi)$ in
the equivalence class of ${\cal V}(\Psi)\simeq {\cal V}(\Psi)+{\cal Q}_{BRST}(\mbox{smth})$,
``uniformly in $\Psi$'', so that (\ref{ExactCovariance}) is true?

The answer to this question is ``no'' in flat space, but ``yes'' in AdS.
It turns out that in $AdS_5\times S^5$ it is possible to choose
the vertex operator to be covariant. 

Let us introduce the notation for the action of the global symmetries 
(compare to (\ref{Collectively})):
\begin{equation}
t^{\overline{n}}_A f(g)  =  
\left.{d\over ds}\right|_{t=0} f(ge^{s t^{\overline{n}}_A})
\end{equation}
\underline{Definition:} The covariant vertex is a 
superfield 
\begin{quote}
\begin{equation}
{\cal V}(\Psi) = 
\lambda^{\alpha}\lambda^{\beta}{\cal V}_{\alpha\beta}(\Psi) +
\lambda^{\alpha}\tilde{\lambda}^{\dot{\beta}}{\cal V}_{\alpha\dot{\beta}}(\Psi) +
\tilde{\lambda}^{\dot{\alpha}}\tilde{\lambda}^{\dot{\beta}} {\cal V}_{\dot{\alpha}\dot{\beta}}(\Psi)
\end{equation}
depending linearly on the state $\Psi$, and such that:
\begin{enumerate}

\item   It is annihilated by ${\cal Q}$:
        \begin{equation}
         (\lambda^{\alpha} {\cal T}_{\alpha} +
          \tilde{\lambda}^{\dot{\alpha}} {\cal T}_{\dot{\alpha}}) {\cal V}(\Psi) =  0
        \end{equation}
        and is not $\cal Q$-exact, and 

\item   the action of the global symmetry on ${\cal V}$ as a function of $g$ agrees with 
        the action of the global symmetry on the space of states:
        \begin{equation}
         t^{\bar{n}}.{\cal V}(\Psi) = {\cal V}(t^{\bar{n}}\Psi)
        \end{equation}
\end{enumerate}
\end{quote}

\section{Taylor series for the vertex.}
\label{sec:TaylorSeries}
Let us study vertex operators for states which are not necessarily
normalizable. In other words, let us forget about the boundary conditions
near the boundary of AdS, and study the supergravity states which are
not necessarily normalizable. Moreover, let us pick a point in $AdS_5\times S^5$
and consider the Taylor expansion of the supergravity fields around this
point. Let us not worry about the convergence of the Taylor series. 
Just study the supergravity equations, BRST cohomology {\it etc.} 
on formal Taylor series. The question of convergence, and the question
of the behaviour at spacial infinity, can be studied later.

For the study of the Taylor series the mathematical notion of the 
{\em coinduced representation} is useful.

\subsection{A review of coinduced representations}

 Let us study 
the supergravity fields around a point in $AdS_5\times S^5$ corresponding to the unit 
${\bf 1}\in G$.
If we do not insist on convergence, then the space of supergravity fields
around a point can be replaced by a more algebraic notion,
the so-called {\em coinduced representation} \cite{Knapp}.

\vspace{10pt}

\noindent
For a representation $V$ of $G_0$, we define  the {\em coinduced representation }
$\mbox{Coind}_{{\bf g}_0}^{\bf g}V$, in the following way: 

\vspace{8pt}
\noindent
\underline{Definition}: 
\begin{quote}
the space of  linear
functions $f$ from the universal enveloping  
algebra  ${\cal U}{\bf g}$ to $V$,
which satisfy the condition of ${\bf g}_0$-invariance:
\[
f(x\xi) = \rho(x)f(\xi) \;\; \mbox{for any}\;\; x\in {\bf g}_0,\; \xi\in {\cal U}{\bf g}
\]
is called the {\em coinduced representation } and denoted
$\mbox{Coind}_{{\bf g}_0}^{\bf g}V$.
\end{quote}

\noindent
The mathematical notation for such functions is:
\begin{equation}
f\in \mbox{Hom}_{{\bf g}_0}({\cal U}{\bf g},V)
\end{equation}
Here ``$\mbox{Hom}(A,B)$'' means the space of linear maps from $A$ to $B$,
and the subindex  ${\bf g}_0$ means ${\bf g}_0$-invariant functions: 
$f(x\xi) - \rho(x)f(\xi) = 0$.

\vspace{5pt}

\noindent The action of ${\bf g}$ on this space is defined by the formula:
\begin{equation}\label{ActionOfG}
x.f(\xi)=f(\xi x)\;,\; \mbox{for}\;\; x\in{\bf g}, \; \xi\in{\cal U}{\bf g}
\end{equation}
To summarize:

\begin{center}
\fbox{
$\mbox{Coind}_{{\bf g}_0}^{\bf g}V = 
\mbox{Hom}_{{\bf g}_0}({\cal U}{\bf g},V)$}
\end{center}

\noindent
We will now explain that $\mbox{Coind}_{{\bf g}_0}^{\bf g}V$ encodes
the Taylor series of various tensor fields on $AdS_5\times S^5$, where $V$ is
the representation of ${\bf g}_0$ corresponding to the type of the tensor.

Let us start with the trivial representation $V={\bf C}$ of ${\bf g}_{\bar{0}}$. 
In this case we should get scalar fields on AdS. The correspondence between
the elements of $\mbox{Hom}_{{\bf g}_0}({\cal U}{\bf g},{\bf C})$ 
and the scalar fields on AdS goes as follows. Given a scalar field
$\phi(g)$, the corresponding element 
$f\in\mbox{Hom}_{{\bf g}_0}({\cal U}{\bf g},{\bf C})$
is given by the formula:
\begin{eqnarray}\label{InducedFromScalars}
f(x_1x_2\cdots x_n) & = & x_1.x_2\ldots x_n.\phi({\bf 1})
\\
\mbox{where} &&
\nonumber
\\
x_1.x_2\ldots x_n.\phi(g) & = &
\left.{\partial\over\partial t_1}\cdots {\partial\over\partial t_n}\right|_{t_1=\ldots=t_n=0}
\phi(ge^{t_1 x_1} e^{t_2 x_2}\cdots e^{t_n x_n})
\nonumber
\end{eqnarray}
In this case the ${\bf g}_0$-invariance condition says that 
$f(x\xi)=0$ for any $x\in {\bf g}_0$, and this is indeed satisfied for
$f$ defined in (\ref{InducedFromScalars}) because $\phi(e^x)=\phi({\bf 1})$
for any $x\in {\bf g}_0$
because $\phi$ is well defined on $G/G_0$.

There is also a map going in the opposite direction. Namely, given 
$f$  a linear function from ${\cal U}{\bf g}$ to ${\bf C}$ we define the
corresponding scalar field $\phi(g)$ as follows:
\begin{equation}
\phi(g)=f(g)
\end{equation}
Note that on the right hand side we treat $g$ as a {\em group element}
\footnote{The $\xi\in {\cal U}{\bf g}$ is called {\em group element}
if it is of the form $\xi=e^x$ for some $x\in {\bf g}$} of ${\cal U}{\bf g}$.

We have just explained why for $V={\bf C}$ the space 
$\mbox{Hom}_{{\bf g}_0}({\cal U}{\bf g},V)$ encodes the Taylor coefficients
of the scalar function on AdS; for general $V$ a similar
construction shows that $\mbox{Hom}_{{\bf g}_0}({\cal U}{\bf g},V)$ encodes
the Taylor coefficients of the tensor field with indices transforming
in the representation $V$ of ${\bf g}_0$.

\subsection{Pure spinors}
\label{sec:PureSpinors}
To describe the linearized SUGRA  in $AdS_5\times S^5$ we need two pure spinor
variables $\lambda_3\in {\bf g}_3$ and $\lambda_1\in {\bf g}_1$
satisfying the constraints:
\begin{equation}\label{PureSpinorConstraints}
\{\lambda_3,\lambda_3\}=\{\lambda_1,\lambda_1\}=0
\end{equation}
We will consider various types of vertex operators, which are homogeneous polynomials
in $\lambda_3$ and $\lambda_1$. Note that (\ref{PureSpinorConstraints}) are
invariant under the action of ${\bf g}_0$. Therefore the polynomials of 
$\lambda_3$ and $\lambda_1$ form a representation of ${\bf g}_0$. We will introduce
the notation for such polynomials:

\definition{We denote ${\cal P}^{(m,n)}$ the space of
polynomials of $\lambda_3$  and $\lambda_1$
which have the degree $m$ 
in $\lambda_3$ and $n$ in $\lambda_1$.}

\noindent
We will define the polynomials of
$\lambda$ by specifying their
coefficients, which are elements of $\Lambda^m {\bf g}'_3 \otimes \Lambda^n {\bf g}'_1$
(see Section \ref{sec:Notations} for notations).
We have to ``discard'' those polynomials which are identically 
zero because of the pure spinor constraints (\ref{PureSpinorConstraints}).

\vspace{10pt}

\noindent
{\small As a trivial example, let us consider the quadratic polynomials
of $\lambda_3$. The coefficients belong to $\Lambda^2 {\bf g}'_3$. 
Let us denote $t_3^{\alpha}$ the basis vectors of ${\bf g}'_3$, such that:
\[
\langle t_3^{\alpha}, t^3_{\beta} \rangle = \delta^{\alpha}_{\beta}
\]
The space $\Lambda^2 {\bf g}'_3$ consists of expressions of the form
$U_{\alpha\beta}t^{\alpha}_3\otimes t^{\beta}_3$ where
$U_{\alpha\beta}=U_{\beta\alpha}$. 
 As an example of the polynomial which is identically
zero, take $f_{\alpha\beta}^{\mu}t_3^{\alpha}\otimes t_3^{\beta}$ where
$f_{\alpha\beta}^{\mu}$ is the structure constants defined by 
$\{ t_{\alpha}^3,t_{\beta}^3 \} = f_{\alpha\beta}^{\mu} t_{\mu}^2$.
Such a polynomial is identically zero because of the pure spinor constraint
(\ref{PureSpinorConstraints}):
\[
f_{\alpha\beta}^{\mu} \lambda_3^{\alpha}\lambda_3^{\beta}=0
\]
}

\vspace{5pt}

\noindent To summarize:
\begin{equation}
{\cal P}^{(m,n)} = (\Lambda^m{\bf g}'_3/(\Lambda^m{\bf g}'_3)_{null})\otimes
(\Lambda^n{\bf g}'_1/(\Lambda^n{\bf g}'_1)_{null})
\end{equation}
where $(\Lambda^m{\bf g}'_3)_{null}$ denotes a subspace of $\Lambda^m{\bf g}'_3$
corresponding to those polynomials on ${\bf g}_3$ which vanish identically on 
$\lambda_3$ because of the pure spinor constraint. 

We will also introduce:
\begin{equation}
{\cal P}^l=\bigoplus\limits_{m+n=l}{\cal P}^{(m,n)}
\end{equation}
Note that ${\cal P}^{(m,n)}$ and ${\cal P}^l$ are representations of
${\bf g}_0$, but not of ${\bf g}$. The construction of coinduced
representation is used to build the representations of ${\bf g}$ from
these spaces.

\subsection{Vertex as an element of $\mbox{Coind}_{{\bf g}_0}^{\bf g}{\cal P}^2$}
In this section we consider the Taylor series of the 
vertex operator
and do not bother about the convergence and the behaviour near the boundary.
Then the vertex operator can be considered an element of the coinduced 
representation:
\begin{equation}
{\cal V}(\Psi)\in \mbox{Coind}_{{\bf g}_0}^{\bf g}{\cal P}^2
\end{equation}
We would like to discuss vertex operators ``uniformly'' for all vectors 
$\Psi\in {\cal H}$. We will therefore introduce the ``universal'' vertex
operator:
\begin{equation}\label{NonCovariantVertex}
{\cal V} \in \mbox{Hom}_{\bf C}({\cal H}, 
\mbox{Coind}_{{\bf g}_0}^{\bf g}{\cal P}^2)
\end{equation}
In other words, we have a linear function on the Hilbert space $\cal H$
which to every  vector $\Psi\in{\cal H}$ associates
the corresponding vertex operator:
\begin{equation}\label{CovariantVertex}
{\cal V}\;:\; \Psi \mapsto {\cal V}(\Psi)\;\;\in \;\;
\mbox{Coind}_{{\bf g}_0}^{\bf g}{\cal P}^2
\end{equation}
Given a state $\Psi\in {\cal H}$ we get ${\cal V}(\Psi)$ --- an element of
$\mbox{Coind}_{{\bf g}_0}^{\bf g}{\cal P}^2$.
This means, by definition, that for every $\Psi$, the object ${\cal V}(\Psi)$
is a linear map from ${\cal U}{\bf g}$ to ${\cal P}^2$ satisfying
the ${\bf g}_0$-invariance condition:
\begin{equation}
{\cal V}(\Psi)(x\xi) = \rho(x)\;{\cal V}(\Psi)(\xi) \;\; \mbox{for any}\;\; x\in {\bf g}_0,\; 
\xi\in {\cal U}{\bf g}
\end{equation}
Given such ${\cal V}(\Psi)$, how do we construct the ``usual'' vertex operator?
As an element of $\mbox{Coind}|_{{\bf g}_0}^{\bf g}{\cal P}^2$ our ${\cal V}(\Psi)$ is
a function of $\xi\in{\cal U}{\bf g}$ with values in ${\cal P}^2$. 
Let us evaluate this function on a group element $\xi=g=e^x$, where $x\in {\bf g}$.
We get ${\cal V}(\Psi)(g)$ --- an element from ${\cal P}^2$, {\it i.e.} a quadratic
polynomial in $\lambda_3$ and $\lambda_1$. The ``usual'' vertex opearator
is just the evaluation of this polynomial:
\begin{equation}\label{UsualVertexOperator}
V_{\Psi}(g,\lambda)=  {\cal V}(\Psi)(g)(\lambda)
\end{equation}

\subsection{Action of the BRST operator}
The BRST complex is:
\begin{equation}
\ldots \stackrel{Q_{BRST}}{\longrightarrow} 
\mbox{Hom}_{\bf C}({\cal H}, 
\mbox{Coind}_{{\bf g}_0}^{\bf g}{\cal P}^n)
\stackrel{Q_{BRST}}{\longrightarrow}
\mbox{Hom}_{\bf C}({\cal H}, 
\mbox{Coind}_{{\bf g}_0}^{\bf g}{\cal P}^{n+1})
\stackrel{Q_{BRST}}{\longrightarrow} \ldots
\end{equation}
The BRST operator acts on the universal vertex ${\cal V}(\Psi)$ in the following
way:
\begin{equation}\label{QonUniversal}
(Q_{BRST}{\cal V})(\Psi)(\xi)(\lambda)=
{\cal V}(\Psi)(\lambda_3\xi + \lambda_1\xi)(\lambda)
\end{equation}
Note that $Q_{BRST} {\cal V}$ is an element of 
$\mbox{Hom}_{\bf C}({\cal H}, 
\mbox{Coind}_{{\bf g}_0}^{\bf g}{\cal P}^3)$.
In terms of the ``usual'' vertex $V_{\Psi}(g,\lambda)$ defined by (\ref{UsualVertexOperator})
we get:
\begin{equation}
Q_{BRST}V_{\Psi}(g,\lambda) = \left.{d\over dt}\right|_{t=0}
V_{\Psi}(e^{t(\lambda_3+\lambda_1)}g,\lambda)
\end{equation}

\subsection{Covariant universal vertex}
\label{sec:CovariantUniversal}
\paragraph     {Statement of covariance}
Note that in Eq. (\ref{NonCovariantVertex}) we use the notation $\mbox{Hom}_{\bf C}$ rather
than $\mbox{Hom}_{\bf g}$. There is no apriori reason why ${\cal V}$ would
respect the action of ${\bf g}$. But in the next section we will see that  
under some conditions on ${\cal H}$, 
it is possible to choose the universal 
vertex operator which does respect the global
symmetry. We will call it the {\em covariant} universal vertex:
\begin{equation}
{\cal V} \in \mbox{Hom}_{\bf g}({\cal H}, 
\mbox{Coind}_{{\bf g}_0}^{\bf g}{\cal P}^2)
\end{equation}
Given Eq. (\ref{ActionOfG}) this implies:
\begin{equation}\label{ActionOfGOnCovariant}
{\cal V}(\Psi)(\xi x)  +  {\cal V}(x \Psi) (\xi) = 0
\end{equation}
\paragraph     {Condition on ${\cal H}$: sufficiently high spin}
The conditions on ${\cal H}$ are the following. Consider ${\cal H}$
as a representation of $so(6)\subset {\bf g}$ --- the symmetry algebra of $S^5$. 
As a representation of $so(6)$, ${\cal H}$ is the direct sum of infinitely many
finite-dimensional representations of $so(6)$. 
We request that the minimal value of the quadratic Casimir of
$so(6)$  on ${\cal H}$ be sufficiently high.

\section{Existence of the covariant vertex}
\label{sec:CohomologicalArgument}
In this Section we will use some facts about the Lie algebra cohomology
which we learned mostly from \cite{Knapp,Fuchs,FeiginFuchs}.
See Chapter 3~\S 6 of \cite{GelfandManin} for a very brief summary.

\subsection{Brief summary}
The physical states correspond to the cohomology of $Q_{BRST}$ at 
the ghost number 2, therefore: 
\begin{equation}
\mbox{Hom}_{\bf g} ({\cal H}, H^2(Q_{BRST}, 
\mbox{Coind}|_{{\bf g}_0}^{\bf g}{\cal P}^{\bullet})) = {\bf C}^{d({\cal H})}
\end{equation}
where $d({\cal H})$ is the multiplicity of ${\cal H}$ (how many times ${\cal H}$
enters in the SUGRA spectrum on $AdS_5\times S^5$). 
We will argue that the second cohomology of the BRST operator
can be calculated using the covariant subcomplex. In other
words,
\begin{equation}\label{EqualsToCovariantH}
\mbox{Hom}_{\bf g} ({\cal H}, H^2(Q_{BRST}, 
\mbox{Coind}|_{{\bf g}_0}^{\bf g}{\cal P}^{\bullet})) 
=
H^2(Q_{BRST}, \mbox{Hom}_{\bf g}({\cal H}, 
\mbox{Coind}_{{\bf g}_0}^{\bf g}{\cal P}^{\bullet}))
\end{equation}
(Notice that 
$\mbox{Hom}_{\bf g}({\cal H}, \mbox{Coind}_{{\bf g}_0}^{\bf g}{\cal P}^{\bullet})$
is the covariant subcomplex.) To prove Eq. (\ref{EqualsToCovariantH}) we
rewrite it in the following form:
\begin{align}
& H^0(\;\;{\bf g}\;\;, \;\;
\mbox{Hom}_{\bf C} (\;{\cal H}\;,\; H^2(Q_{BRST}, 
  \mbox{Coind}|_{{\bf g}_0}^{\bf g}{\cal P}^{\bullet})\;)\;\;)
= 
\nonumber
\\ 
 =\; & H^2(\;\;Q_{BRST}\;\; , \;\; 
  H^0(\;{\bf g} \;,\; \mbox{Hom}_{\bf C}({\cal H}, 
  \mbox{Coind}_{{\bf g}_0}^{\bf g}{\cal P}^{\bullet})\;)\;\;)
\label{HPermuted}
\end{align}
Here we have used the fact that for any representation $L$
of the Lie algebra ${\bf g}$ the zeroth cohomology group
$H^0({\bf g},L)$ equals the space of invariants $\mbox{Inv}_{\bf g}L$.
In particular,  for  two representations ${\cal A}$ and ${\cal B}$,
$ H^0(\;{\bf g}\;,\; \mbox{Hom}_{\bf C}({\cal A},{\cal B}) \;) = 
 \mbox{Hom}_{\bf g}({\cal A},{\cal B}) $.

The idea of the proof of (\ref{HPermuted}) is to note that
the left and the right hand side of (\ref{HPermuted}) are
two different  second approximations to calculating the cohomology
of the ``total'' differential $Q_{BRST} + Q_{Lie}$. Therefore
the equality of the left hand side and the right hand side follows
if we prove that the second approximation is actually exact.
To prove that we will need several vanishing theorems. These vanishing
theorems essentially follow from the fact that as a representation
of $so(6)$ (the rotations of $S^5$) ${\cal H}$ is a direct sum of finite-dimensional 
representations. 
 This can be seen from the explicit description of the
supergravity solutions in \cite{KimRomansNieuwenhuizen}.

\subsection{Bicomplex and spectral sequence}
\label{sec:SpectralSequence}

Let us start by fixing  some universal vertex (not necessarily covariant):
\begin{equation}
{\cal V}\;:\; {\cal H}\to \mbox{Coind}_{{\bf g}_0}^{\bf g} {\cal P}^2
\end{equation}
At this point we do not require that this vertex is covariant; it is apriori an
element of $\mbox{Hom}_{\bf C}({\cal H},\mbox{Coind}_{{\bf g}_0}^{\bf g} {\cal
  P}^2)$ rather than $\mbox{Hom}_{\bf g}({\cal H},\mbox{Coind}_{{\bf g}_0}^{\bf g}
{\cal P}^2)$.  
We will introduce the Lie algebra BRST operator of ${\bf g}$. For
each generator $t_i$ of ${\bf g}$ we introduce the corresponding ghost $c^i$, and
define:
\begin{equation}
Q_{Lie}=c^it_i - {1\over 2} f_{ij}^k c^ic^j {\partial\over\partial c^k}
\end{equation}
We will consider the action of $Q_{Lie}$ on expressions polynomial in $c^i$.
The polynomials of $c^i$ are specified by their coefficients; in degree $l$
the coefficients live in $\Lambda^l {\bf g}^*$.
Therefore $Q_{Lie}$ acts on the vertex operator as follows:
\begin{equation}
\mbox{Hom}_{\bf C}({\cal H},\mbox{Coind}_{{\bf g}_0}^{\bf g} {\cal P}^2)
\stackrel{Q_{Lie}}{\longrightarrow}
\mbox{Hom}_{\bf C}({\cal H},\mbox{Coind}_{{\bf g}_0}^{\bf g} {\cal P}^2)\otimes 
{\bf g}
\end{equation}
We will consider the bicomplex with the differential $Q_{tot}$:
\begin{equation}
Q_{tot} = Q_{BRST} + Q_{Lie}
\end{equation}
To prove the existence of the covariant vertex we will consider
the spectral sequence computing the cohomology of this bicomplex. 
There are two ways to construct the spectral sequence. One can
first calculate the cohomology of $Q_{BRST}$ and then consider $Q_{Lie}$
as a perturbation.
The other way is to first calculate the cohomology of $Q_{Lie}$ and
then consider $Q_{BRST}$ as a perturbation. These two ways of calculating
the cohomology of $Q_{tot}$ should give the same result. We will see that
this implies the existence of the covariant vertex.
We will now consider the two methods in turn.

\paragraph     {First $Q_{BRST}$ then $Q_{Lie}$} 
The first term of the spectral sequence has:
\begin{equation}
E_1^{p,q} = H_{Q_{BRST}}^q(\mbox{Hom}_{\bf C}({\cal H}, 
\mbox{Coind}_{{\bf g}_0}^{\bf g} {\cal P}^{\bullet}))\otimes \Lambda^p {\bf g}'
\;\;,\;\;\;
d_1:\; E_1^{p,q} \to E_1^{p+1,q}
\end{equation}
where $\Lambda^p {\bf g}'$ stands for the  $c$-ghosts; 
an element of $E_1^{p,q}$ is schematically $\lambda^q c^p$. The differential in the first
term is $d_1 = Q_{Lie}$.
The second term is:
\begin{equation}
E_2^{p,q} = H^p( \;\; {\bf g}, \; H_{Q_{BRST}}^q(\mbox{Hom}_{\bf C}({\cal H}, 
\mbox{Coind}_{{\bf g}_0}^{\bf g} {\cal P}^{\bullet}))\;\; )\;,\;\;\;
d_2\;: \; E_2^{p,q} \to E_2^{p+2,q-1}
\end{equation}
The higher differentials are of the type $d_r:\; E_r^{p,q}\to E_r^{p+r,q+1-r}$.

\paragraph     {First $Q_{Lie}$ then $Q_{BRST}$} 
The first term is:
\begin{equation}
\tilde{E}_1^{p,q} = H^p(\; {\bf g}, \;
\mbox{Hom}_{\bf C}({\cal H}, \mbox{Coind}_{{\bf g}_0}^{\bf g}{\cal P}^q)\;)\;,
\; \tilde{d}_1\;:\; \tilde{E}_1^{p,q} \to \tilde{E}_1^{p,q+1}
\end{equation}
where $\tilde{d}_1 = Q_{BRST}$. The higher differentials are of the type
$\tilde{d}_r: \; \tilde{E}_r^{p,q}\to \tilde{E}_r^{p+1-r,q+r}$.

\paragraph     {Existence of the covariant vertex}
\begin{figure}[th]
\centerline{
\includegraphics[width=2.0in]{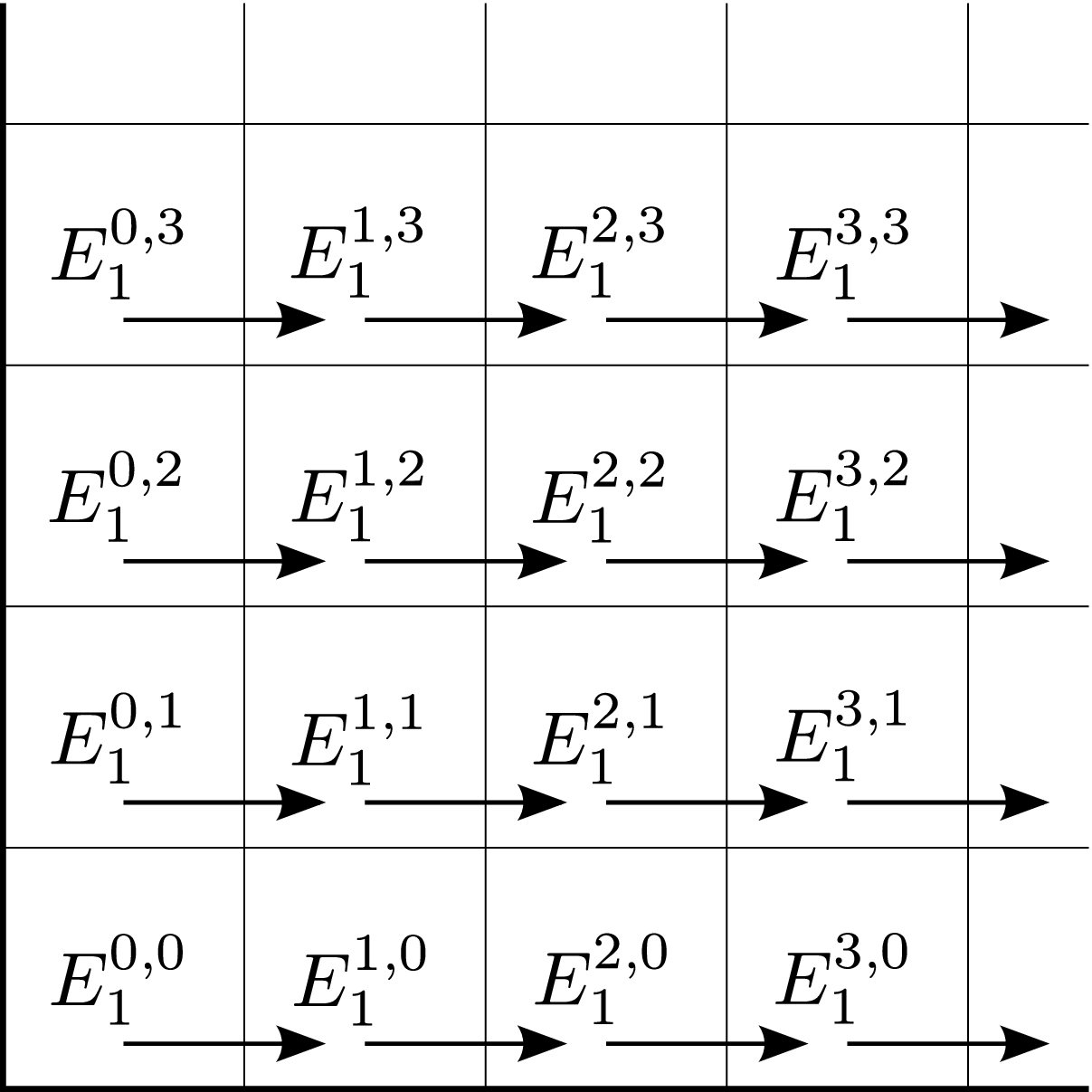}
\hspace{20pt}
\includegraphics[width=2.0in]{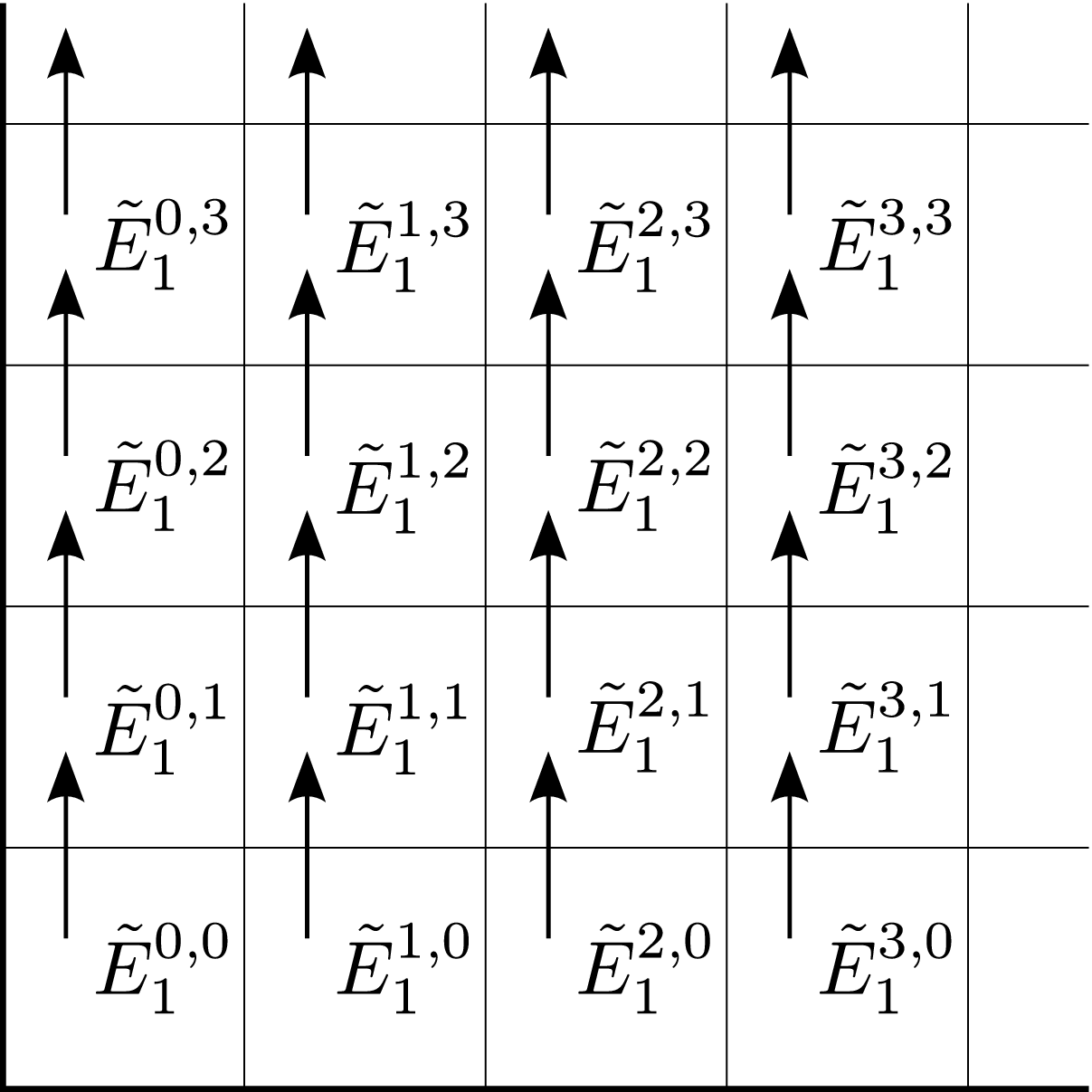}
}
\caption{\label{fig:FirstPage}\small
The first page of $E$ and $\tilde{E}$; arrows denote $d_1$ and
$\tilde{d}_1$. The Lie algebra ghost number (the number of $c$'s)
increases in the horizonthal direction, while the BRST ghost number
(the number of $\lambda_3$ plus the number of $\lambda_1$) in the
vertical direction.
}
\end{figure}

\begin{figure}[th]
\centerline{
\includegraphics[width=2.0in]{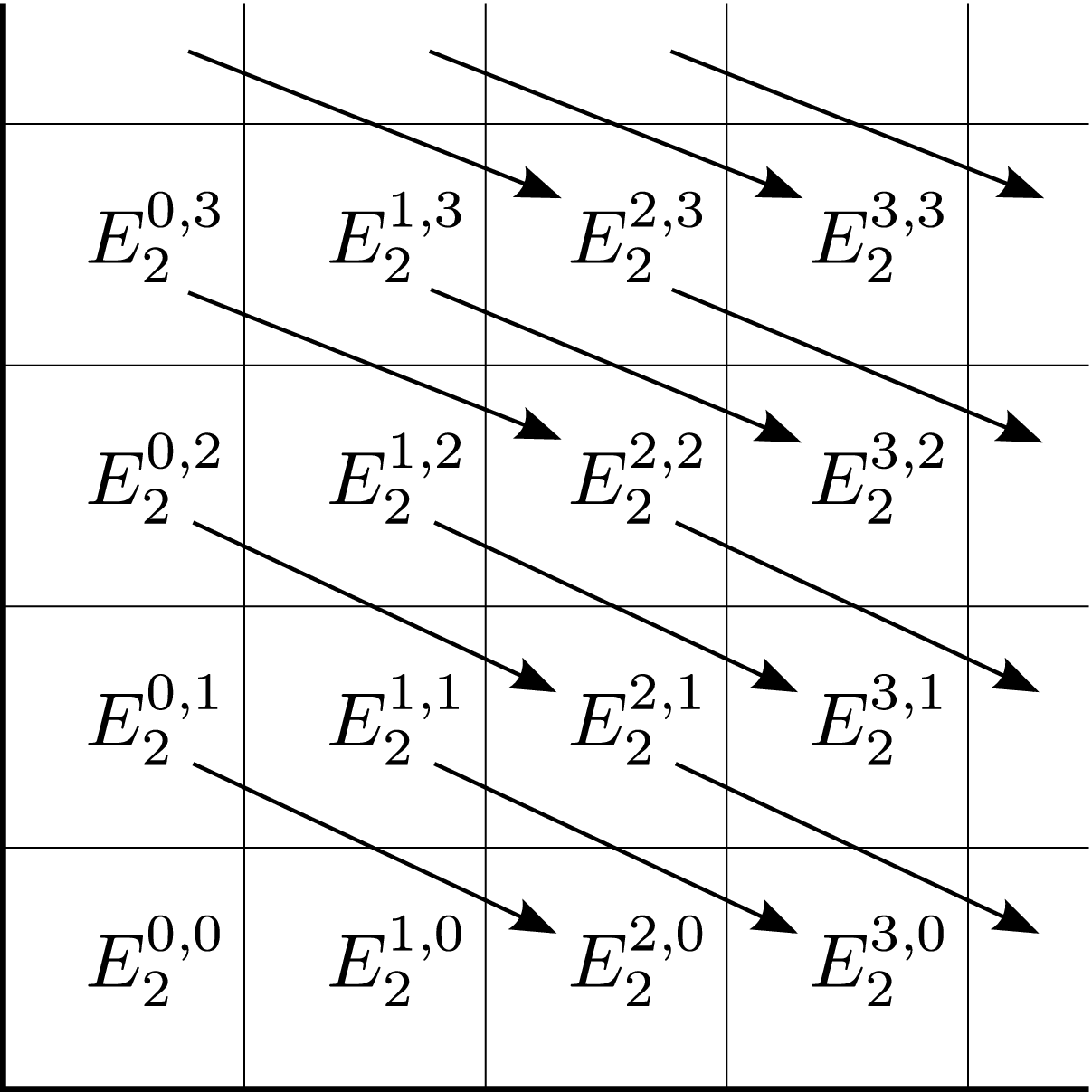}
\hspace{20pt}
\includegraphics[width=2.0in]{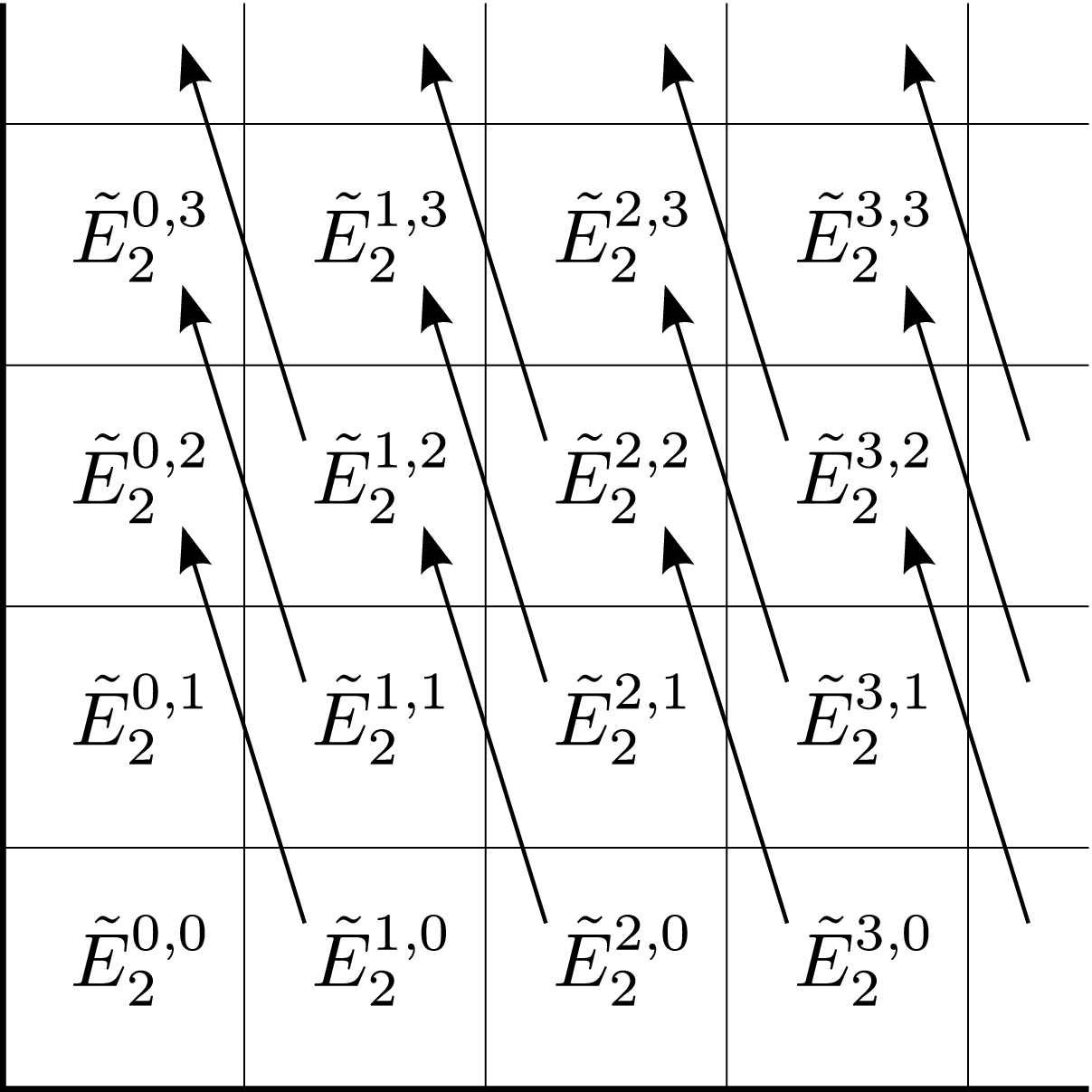}
}
\caption{\label{fig:SecondPage}\small
The second page of $E$ and $\tilde{E}$; arrows denote $d_2$ and
$\tilde{d}_2$.
}
\end{figure}

\begin{figure}[th]
\centerline{
\includegraphics[width=2.0in]{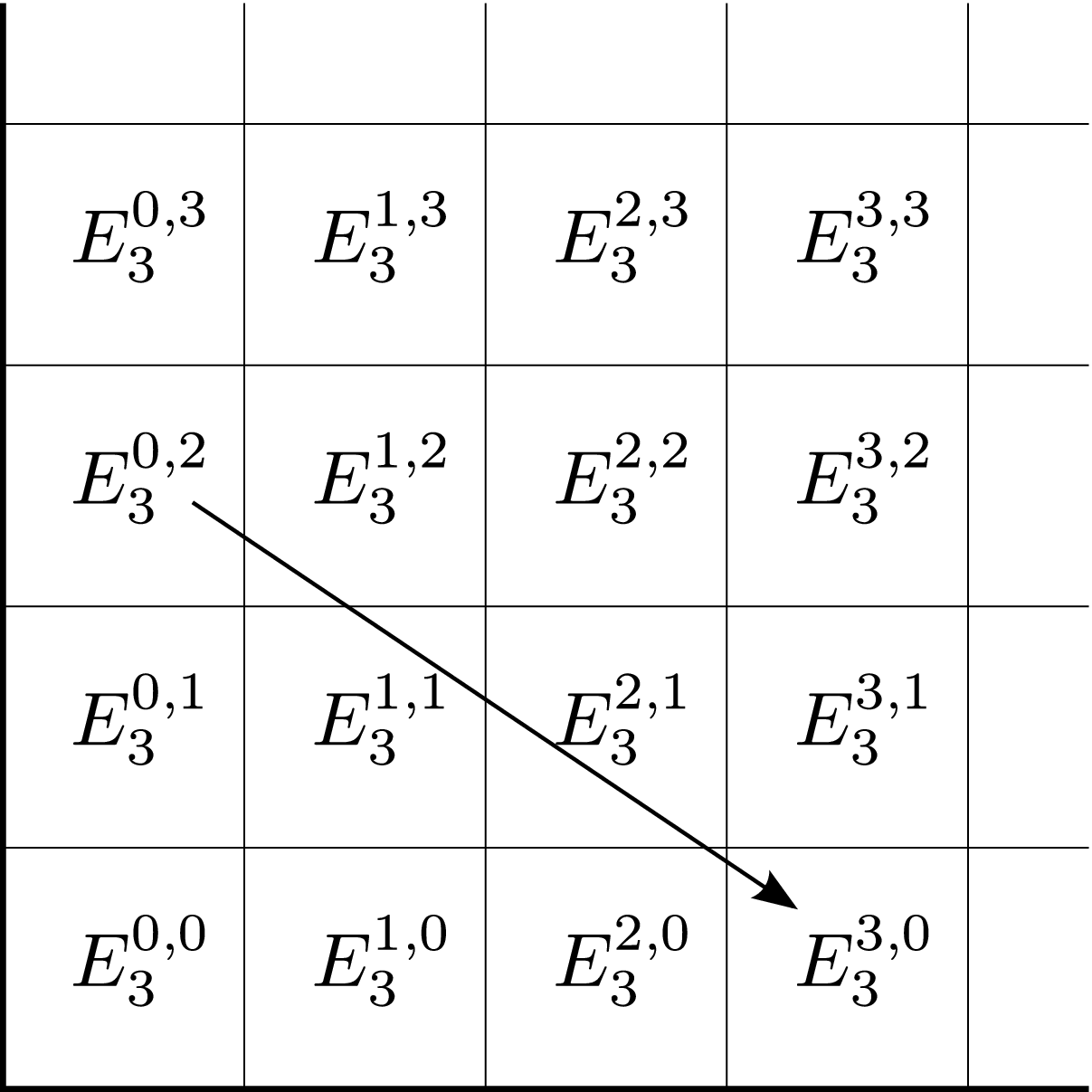}
}
\caption{\label{fig:Transgression}\small
The differential $d_3$ of $E_3$.
}
\end{figure}
First of all, we want to show that  $E^{0,2}_{\infty} = E^{0,2}_1$. The first observation
is that by definition $d_1: E_1^{0,2}\to E_1^{1,2}$ is zero. This is
because the vertex is {\em covariant up to BRST-exact correction}
(see Eq. (\ref{UpToBRSTExact})). Therefore $E_1^{0,2}= E_2^{0,2}$.
Also, we will show (for ${\cal H}$
with large enough spin) that $E_2^{2,1} = E_2^{3,0} = 0$. 
This implies that $E^{0,2}_{\infty} = E^{0,2}_1$. 

Then we remember the relation
between $E^{0,2}_{\infty}$ and $H^2(Q_{tot})$, which is the following. The space $H^2(Q_{tot})=E^2$
has a filtration, corresponding to the number of the $c$-ghosts.
Namely, $F^pE^2$ consists of expressions containing at least $p$
$c$-ghosts. Then $E_{\infty}^{0,2} = E^2/F^1E^2 = H^2(Q_{tot})/F^1H^2(Q_{tot})$. 
To summarize:
\begin{eqnarray}\label{H2QtotFromE}
E_1^{0,2} & = & [\mbox{unintegrated vertices}]
\\[3pt]
E_1^{0,2} & = & H^2(Q_{tot}) / F^1H^2(Q_{tot})
\label{E1IsFactorspace}
\end{eqnarray}
On the other hand we will show that 
$\tilde{E}_1^{1,1} = \tilde{E}_1^{2,0} = 0$ (see Eqs. (\ref{ShapiroForH2}), (\ref{ShapiroForH1}))
and also that $\tilde{E}_1^{1,0}=0$ (similar to (\ref{ShapiroForH2})). This implies that:
\begin{equation}\label{H2QtotFromEtilde}
H^2(Q_{tot}) = 
{\mbox{Ker}\; \tilde{d}_1:\; \tilde{E}_1^{0,2}\to \tilde{E}_1^{0,3} 
\over 
 \mbox{Im}\;  \tilde{d}_1:\; \tilde{E}_1^{0,1}\to \tilde{E}_1^{0,2}}
\end{equation}
We are now ready to prove the existence of the covariant vertex.
Notice that $\tilde{E}_1^{0,q} = H^0(\; {\bf g}, \;
\mbox{Hom}_{\bf C}({\cal H}, \mbox{Coind}_{{\bf g}_0}^{\bf g}{\cal P}^q)\;)$
is the space of functions $f_{\Psi}(x,\theta,\lambda)$, parametrized
by $\Psi\in {\cal H}$, transforming covariantly under ${\bf g}$. 
This means that $\tilde{E}_1^{0,\bullet}$ is the {\em covariant 
subcomplex} of the BRST complex. (The subspace consisting of the
covariant expressions.) And Eq. (\ref{H2QtotFromEtilde}) shows
that:
\begin{equation}\label{H2QtotIsCovariant}
H^2(Q_{tot}) \mbox{ is the second cohomology of the covariant subcomplex}
\end{equation}
Now the comparison of
 (\ref{H2QtotFromE}), (\ref{E1IsFactorspace}) and (\ref{H2QtotIsCovariant}) shows that the 
cohomology of $Q_{BRST}$ can be calculated using the covariant subcomplex.
In fact, if the representation ${\cal H}$ has large enough momentum in $S^5$,
then $F^1E^2$ is zero (because already $E^{1,1}_2$ and $E^{2,0}_2$ are zero). This means
that the factor space on the right hand side of (\ref{E1IsFactorspace})
is just $H^2(Q_{tot})$.

This means that there is a covariant choice of the vertex. In the rest
of this section we will prove the required vanishing theorems and
explain explicitly how the non-covariant vertex can be modified into
the covariant one.

\paragraph     {Gauge transformations}
It is also true that $\tilde{E}^{1,0}_1 = 0$, and therefore $\tilde{E}^{1,0}_2 = 0$. This is proven
similarly to (\ref{ShapiroForH2}). This implies that $\tilde{d}_2:\; \tilde{E}^{1,0}_2 \to \tilde{E}^{0,2}_2$ is zero.
This means that when considering the gauge transformations of the covariant
vertices it is enough to consider the gauge transformations with the 
covariant parameters; if a covariant vertex is BRST trivial, then it
is a BRST variation of a covariant expression.

\subsection{The descent}
\label{sec:Descent}

Since the vertex transforms covariantly up to BRST-exact terms, we must have:
\begin{equation}
Q_{Lie}{\cal V}= Q_{BRST} {\cal W} 
\end{equation}
where
\begin{equation}
{\cal W}\in 
\mbox{Hom}_{\bf C}({\cal H},\mbox{Coind}_{{\bf g}_0}^{\bf g} {\cal P}^1)
\otimes  {\bf g}
\end{equation}
Note that $Q_{Lie}{\cal W}$ is $Q_{BRST}$-closed and has ghost number 1:
\begin{equation}
Q_{Lie}{\cal W}\in 
\mbox{Hom}_{\bf C}({\cal H},\mbox{Coind}_{{\bf g}_0}^{\bf g} {\cal P}^1)
\otimes  \Lambda^2{\bf g}
\end{equation}
We want to argue that there exists ${\cal U}$ 
such that $Q_{Lie}{\cal W}=Q_{BRST}{\cal U}$.
More precisely: note that we are free to add to ${\cal W}$ something in the
kernel of $Q_{BRST}$; we want to prove that it is
possible to use this freedom and choose ${\cal W}$ so that
there exists ${\cal U}$ 
such that $Q_{Lie}{\cal W}=Q_{BRST}{\cal U}$.
An obstacle to this would be a nonzero $d_2 {\cal V}$ where
\begin{eqnarray}
& d_2\;:\;\; & 
H^0({\bf g}, \;
    H^2_{Q_{BRST}}(\mbox{Hom}_{\bf C}({\cal H},
                                      \mbox{Coind}_{{\bf g}_0}^{\bf g}{\cal P}^{\bullet})))
\rightarrow
\nonumber
\\
&& \rightarrow 
H^2({\bf g}, \;
    H^1_{Q_{BRST}}(\mbox{Hom}_{\bf C}({\cal H},
                                      \mbox{Coind}_{{\bf g}_0}^{\bf g}{\cal P}^{\bullet})))
\label{CheckerMove}
\end{eqnarray}
We want to argue that the space $H^2({\bf g}, \;
    H^1_{Q_{BRST}}(\mbox{Hom}_{\bf C}({\cal H},
                                      \mbox{Coind}_{{\bf g}_0}^{\bf g}{\bf C})))$
on the right hand side
is zero.
Note that the BRST cohomology in ghost number 1 corresponds to local conserved
charges. 
But the only conserved charges are the global symmetries $psu(2,2|4)$, and 
those transform in the adjoint representation\footnote{We calculate
the {\em covariant} cohomology of $H^1(Q_{BRST}, \mbox{Coind}_{{\bf g}_0}^{\bf g}{\bf C})$ at the ghost number 1 in  Appendix \ref{sec:GhostNumberOne}.
However we did not calculate this cohomology without the assumption of
covariance. But we know from physics that every cohomology class at the ghost
number 1 corresponds to a local conserved current in the pure spinor sigma
model. And we know the classification of the local conserved currents,
they transform in the adjoint of ${\bf g}$.} of ${\bf g}$. This means that on the 
right hand side of (\ref{CheckerMove}) we have:
\begin{equation}\label{FromLocalConservedCharges}
H^2({\bf g}, \;
    \mbox{Hom}_{\bf C}({\cal H}, {\bf g}) )
\end{equation}
\begin{figure}[th]
\centerline{\includegraphics[width=4.9in]{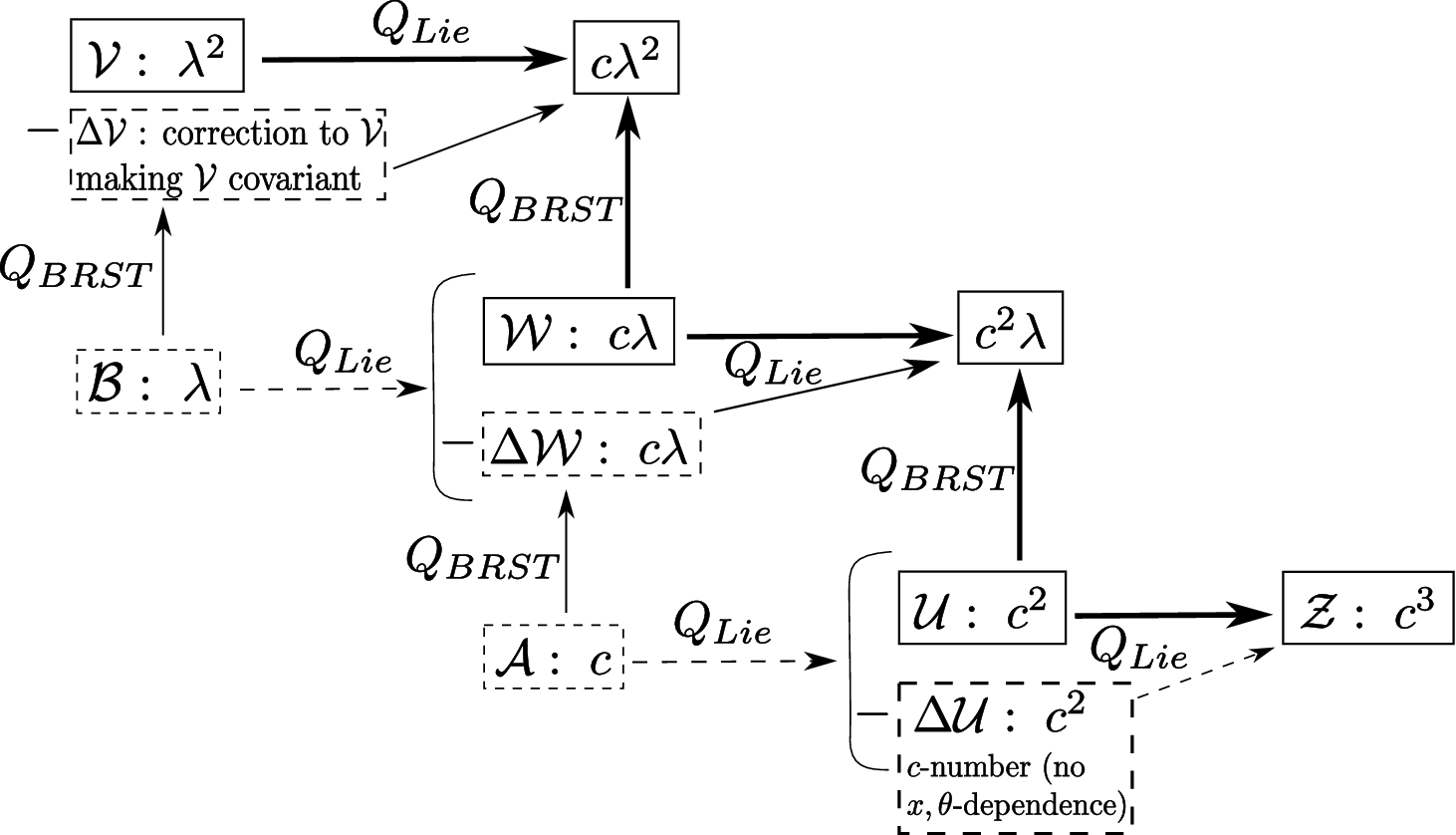}}
\caption{\label{fig:Adjustment}\small
Adjustment of the vertex operator. 
}
\end{figure}
This cohomology group is zero. 
Indeed, we can compute it using the
Serre-Hochschild spectral sequence of ${\bf g}_{even}\subset {\bf g}$. 
Already the first term of this spectral sequence consists of the
following spaces, which are all zero:
\begin{eqnarray}
&& 
\mbox{Hom}_{{\bf g}_{even}}(\Lambda^2 {\bf g}_{odd}, 
\mbox{Hom}_{\bf C}({\cal H}, {\bf g}))\;,\;\;
H^1({\bf g}_{even}, 
\mbox{Hom}_{\bf C}({\bf g}_{odd}\otimes{\cal H}, {\bf g}))\;,
\nonumber
\\
&& H^2({\bf g}_{even}, \mbox{Hom}_{\bf C}({\cal H}, {\bf g}))
\end{eqnarray}
Note that ${\bf g}_{even}={\bf g}_A\oplus {\bf g}_S$ where
${\bf g}_A=so(2,4)$ and ${\bf g}_S=so(6)$.
Consider the corresponding Casimir operators $\Delta_A$ and $\Delta_S$.
For the cohomology to be nonzero, we need both of them zero,
but $\Delta_S$ is positive definite at least for ${\cal H}$ with large enough
momenta. (Note also that ${\cal H}$ is an infinite dimensional
irreducible representation of ${\bf g}$, so there are no invariants
in its tensor product with powers of ${\bf g}$.)

Therefore $Q_{Lie}{\cal W}=Q_{BRST}{\cal U}$ for some ${\cal U}$.
In other words $d_2{\cal V}$ is zero, and we can proceed with
computing $d_3$.

Consider ${\cal Z}=Q_{Lie}{\cal U}$.
Note that ${\cal Z}$ has zero pure spinor ghost number, and $Q_{BRST}{\cal Z}=0$.
Since ${\cal Z}$ is of ghost number 0, 
this implies that ${\cal Z}$ is a constant; it does not contain any $x$ or
$\theta$. Also, we could have added a constant to ${\cal U}$ without affecting
$Q_{BRST}{\cal U}$; therefore ${\cal Z}$ by itself is not very well defined
by our construction. What is well defined is ${\cal Z}$ modulo the image of
$Q_{Lie}$:
\begin{equation}
[{\cal Z}] \in H^3_{Q_{Lie}}({\cal H}'\otimes \Lambda^{\bullet}{\bf g})
= H^3({\bf g},{\cal H}')
\end{equation}

\subsection{The ascent}
But the Lie algebra cohomology group $H^3({\bf g},{\cal H}')$ is zero: 
\begin{equation}\label{H3Zero}
H^3({\bf g},{\cal H}') = 0
\end{equation}
One can see that it is zero from the Serre-Hochschild spectral 
sequence corresponding to ${\bf g}_{even}\subset {\bf g}$. Already the first term
of this spectral sequence consists of the following spaces, which are all zero:
\begin{eqnarray}
&& Hom_{{\bf g}_{even}}(\Lambda^3{\bf g}_{odd},{\cal H}')  
=
H^1({\bf g}_{even} , Hom_{\bf C}(\Lambda^2{\bf g}_{odd},{\cal H}') ) 
=
\nonumber
\\
&=& 
H^2({\bf g}_{even} , \mbox{Hom}_{\bf C}({\bf g}_{odd},{\cal H}') ) 
=
H^3({\bf g}_{even} , {\cal H}')
=0
\end{eqnarray}
The vanishing of these cohomologies can be proven as follows. Note that
${\bf g}_{even}$ splits into ${\bf g}_A=so(2,4)$ and ${\bf g}_S=so(6)$.
For the cohomology to be nontrivial, both ${\Delta}_A$ and ${\Delta}_S$ should be
zero. But $-\Delta_S$ is positive definite. 

Therefore we can remove ${\cal Z}$ by modifying ${\cal U}$, adding
to ${\cal U}$ a constant term $-\Delta {\cal U}$ so that the modified 
${\cal U}-\Delta {\cal U}$ has
$Q_{Lie}({\cal U}-\Delta{\cal U})=0$. (Note that adding the constant term does not
change the image of $\cal U$ under $Q_{BRST}$.) 
Is it possible to find such ${\cal U}'$
that ${\cal U}-\Delta{\cal U}=Q_{Lie}{\cal U}'$? The answer is ``yes'', because
\begin{equation}
H^2({\bf g}, 
\mbox{Hom}_{\bf C}({\cal H}, \mbox{Coind}_{{\bf g}_0}^{\bf g}{\bf C}))=0
\end{equation}
This can be proven using the Shapiro's lemma (Proposition 6.8 and
Theorem 6.9 from \cite{Knapp}; see Appendix \ref{sec:Shapiro} for a review):
\begin{eqnarray}
H^2({\bf g}, \mbox{Hom}_{\bf C}({\cal H},\mbox{Coind}|_{{\bf g}_0}^{\bf g}{\bf C})) =
\nonumber
\\
=\mbox{Ext}^2_{\bf g}({\cal H},\mbox{Coind}|_{{\bf g}_0}^{\bf g}{\bf C})=
\nonumber
\\
\label{ShapiroForH2}
=\mbox{Ext}^2_{{\bf g}_0}({\cal H}|_{{\bf g}_0},{\bf C})=
\\
=H^2({\bf g}_0,{\cal H}'|_{{\bf g}_0})
\nonumber
\end{eqnarray}
Note that ${\bf g}_0=so(1,4)\oplus so(5)$. 
We want to prove that $H^2({\bf g}_0,{\cal H}'|_{{\bf g}_0}) = 0$.
The space ${\cal H}'$ consists of functionals on the space of states.
Since we work in the vicinity of the fixed point $x_0\in AdS_5\times S^5$
our ${\cal H}'$ is generated by the values of various supergravity
fields at the point $x_0$. For example the Ramond-Ramond
field strength $H_{ijk}(x_0)$  and its derivatives. Under the action
of $so(1,4)\oplus so(5)$ this space splits into infinitely many
finite-dimensional representations. For example $\partial_i H_{kjl}(x_0)$
lives in $(\mbox{Vect}\otimes \Lambda^3\mbox{Vect})_0$ where
Vect is is the vector representation of $so(1,4)\oplus so(5)$
and index $0$ means that the contraction $g^{ij}\partial_i H_{jkl}$ 
is zero\footnote{this is an over-simplification; 
in fact one has to add also the expression
of the form $g^{pp'}g^{qq'}F_{iklpq}H_{jp'q'}$, for the contraction
to be zero; the Ramond-Ramond 5-form $F_{iklpq}$ is nonzero in the
AdS background.}. It follows from the general theory of Lie algebra
cohomology that $H^2$ of $so(1,4)\oplus so(5)$ with coefficients
in any finite-dimensional representation is zero.

These arguments imply that ${\cal U}'$ is in the image of $Q_{Lie}$.
We can modify ${\cal W}$ by adding to it: 
\begin{equation}
\Delta {\cal W} = Q_{BRST}Q_{Lie}^{-1}{\cal U}'
\end{equation}
Then we have:
\begin{equation}
Q_{Lie}({\cal W} + \Delta {\cal W}) = 0
\end{equation}
Now we use:
\begin{eqnarray}
H^1({\bf g},\mbox{Hom}_{\bf C}({\cal H},\mbox{Coind}_{{\bf g}_0}^{\bf g} {\bf g}_{odd})) =
\nonumber
\\
\label{ShapiroForH1}
= H^1({\bf g}_0,\;{\cal H}'|_{{\bf g}_0}\otimes_{\bf C} {\bf g}_{odd}) =0
\end{eqnarray}
Therefore ${\cal W}+\Delta {\cal W}$ is in the image of $Q_{Lie}$.

\vspace{5pt}
\noindent
Now the {\em modified vertex:}
\begin{equation}
{\cal V}+Q_{BRST}Q_{Lie}^{-1}({\cal W} + \Delta {\cal W})
\end{equation}
is covariant.

\vspace{5pt}
\noindent
Our  procedure could perhaps be summarized as follows:
\begin{eqnarray}
{\cal V}_{covariant} & = & {\cal V} +
Q_{BRST}Q_{Lie}^{-1}({\cal W} + Q_{BRST}Q_{Lie}^{-1}({\cal U}-Q_{Lie}^{-1}{\cal Z}))
\\
\mbox{where} &&
\nonumber
\\
{\cal W} & = & Q_{BRST}^{-1}Q_{Lie} {\cal V}
\nonumber
\\
{\cal U} & = & Q_{BRST}^{-1}Q_{Lie} {\cal W}
\nonumber
\\
{\cal Z} & = & Q_{Lie} {\cal U} \;\;
\mbox{\small (does not depend on $x,\theta$)}
\nonumber
\end{eqnarray}

\section{How the descent procedure works  in the flat space limit.}
\label{sec:FlatSpaceLimit}
In flat space it  is impossible to choose a covariant vertex, because of the
nontrivial cohomology
\begin{equation}
{\cal Z} \in H^3(\mbox{super-Poincare algebra}, {\bf C})
\end{equation}
which represents the NSNS 3-form field strength. 

But one can satisfy a weaker covariance condition. Note that in flat space
the generators of the Lorentz subalgebra $so(1,9)$ of the Poincare algebra
can not be obtained as commutators of other generators. Therefore it is
consistent to require the covariance under all translations and supersymmetries,
but only some rotations. In particular, it turns out that we can choose
a vertex covariant under:
\begin{equation}\label{PSmall}
{\bf SP}_{small} = 
\{\mbox{translations, supersymmetries, and } so(1,4)\oplus so(5)\subset so(1,9)\}
\end{equation}
This is a subalgebra of the super-Poincare algebra:
\begin{equation}
{\bf SP}_{small} \subset {\bf SP}
\end{equation}
corresponding to the split of the space-time:
\begin{equation}
{\bf R}^{1+9} = {\bf R}_A^{1+4}\times {\bf R}_S^{5}
\end{equation}
We will say that the ten spacetime directions split into $1+4$ A-directions
and $5$ S-directions (the letters A and S stand for the AdS and the sphere). 

Let us now explain how the  diagramm of Fig. \ref{fig:Adjustment} works in flat space.

\subsection{Maxwell field}
Instead of considering Fig. \ref{fig:Adjustment} literally let  us study the similar
diagramm for the supersymmetric Maxwell field (rather than supergravity). 
This is a toy model; the  supersymmetric
Maxwell field in flat space is  ``one half of the supergravity field''.
The ``usual'' (non-covariant) vertex operator is of the form:
\begin{equation}\label{VertexOperator}
V(x,\theta) = (\lambda\Gamma^{\mu}\theta)a_{\mu} 
+(\lambda\Gamma^{\mu}\theta)(\psi\Gamma^{\mu}\theta)  
-{1\over 4} (\theta \Gamma^{\mu\nu\rho} \theta)(\lambda \Gamma^{\rho} \theta)
\partial_{[\mu}a_{\nu]} + \ldots 
\end{equation}
where $a_{\mu}=a_{\mu}(x)$ and $\psi=\psi(x)$ are the  vector potential and the
photino. 

\subsubsection{Action of the  Poincare algebra}
Let us first try to understand if it is possible to choose the vertex covariant
under the even Poincare algebra.  The vertex operator (\ref{VertexOperator})
involves the gauge field $a_{\mu}$. Because of the gauge invariance the gauge
field is not in one to one correspondence with the physical states. 
The physical states are described by 
$f_{\mu\nu}=\partial_{\mu}a_{\nu}-\partial_{\nu}a_{\mu}$, not by $a_{\mu}$.
To describe $a_{\mu}$ in terms of $f_{\mu\nu}$, let us break the translational
symmetries by choosing a point $0$ in space-time. Then we can write, in
the vicinity of the chosen point:
\begin{equation}\label{VicinityGauge}
a=\iota_E {1\over {\cal L}_E} f 
\end{equation}
where $E=x^{\mu}{\partial\over \partial x^{\mu}}$ and $\iota_E$ and ${\cal L}_E$
are the corresponding $\iota$ and Lie derivative. For example,
${\cal L}_E (xdx) = 2xdx$ and $\iota_E dx = x$.
Note that Eq. (\ref{VicinityGauge}) is one particular way to choose 
a vector potential with the field strength $f$.

Let us therefore replace $a$ with $\iota_E {1\over {\cal L}_E} f$.
This breaks the translation symmetry, since the gauge (\ref{VicinityGauge})
depended on a choice of point $x=0$. Can we restore the translational symmetry?
Let us introduce the operator $Q_{Lie}$ 
acting on
the physical vertex operators in the following way:
\begin{equation}
Q_{Lie} V^I = c^{\mu} ( t_{\mu}.V^I - t_{\mu J}^I V^J )
\end{equation}
Here the index $I$ runs over an infinite set enumerating 
the basis vectors of ${\cal H}$, and $t_{\mu}$ are the generators
of translations $\partial\over\partial x^{\mu}$. Fermionic parameters
$c^{\mu}$ are the Lie-algebraic ghosts of the translation algebra.
This operator $Q_{Lie}$ measures the deviation of the vertex operator 
from transforming covariantly under
the action of the global shift. We observe that $Q_{Lie}V$ is $d$ of
something:
\begin{eqnarray}
Q_{Lie} \left(\iota_E {1\over {\cal L}_E} f\right)  & = &
\left[{\cal L}_c \;,\; \iota_E {1\over {\cal L}_E} \right] f
=
\left( \iota_c {1\over {\cal L}_E} - 
\iota_E {1\over {\cal L}_E} {\cal L}_c {1\over {\cal L}_E}\right) f =
\nonumber
\\
&=& 
d\left( {1\over {\cal L}_E({\cal L}_E+1)} \iota_E\iota_c f\right)
\label{DescentDeRham}
\end{eqnarray}
Let us calculate $Q_{Lie}$ of this ``something'':
\begin{eqnarray}
&& Q_{Lie}\left( {1\over {\cal L}_E({\cal L}_E+1)} \iota_E\iota_c f\right)
=
\left\{ {\cal L}_c \;,\; 
{1\over {\cal L}_E({\cal L}_E+1)} \iota_E\iota_c \right\} f
=
\nonumber
\\
&&=
{1\over ({\cal L}_E+1)({\cal L}_E+2)}{\cal L}_c\; \iota_E\; \iota_c  \; f+
{1\over  {\cal L}_E ({\cal L}_E+1)}\; \iota_E\; \iota_c\;  {\cal L}_c\;  f
=
\nonumber
\\
&&=
{1\over ({\cal L}_E+1)({\cal L}_E+2)} \iota_c^2  \; f
+
{2\over {\cal L}_E ( {\cal L}_E+1 )( {\cal L}_E+2 )}
\iota_E\iota_c {\cal L}_c f
\label{EqualsConstant}
\end{eqnarray}
Expanding $f$ in Taylor series around $x=0$ and taking into account that
$df=0$ we can see that (\ref{EqualsConstant}) is equal to:
\begin{equation}\label{Descent}
{1\over 2} \iota_c^2 f(0)
\end{equation}
We should stress that (\ref{EqualsConstant}) is equal to the {\em constant} 
(independent of $x$) expression (\ref{Descent}). In other words, the only
term in the Taylor expansion of (\ref{EqualsConstant}) around the point 
$x=0$ is the constant term. One can see it, for example, because the Lie
derivative ${\cal L}_E$ of (\ref{EqualsConstant}) vanishes. This can be
seen from the identity 
${\cal L}_E \iota_c^2 f + 2\iota_E\iota_c {\cal L}_c f = 0$ which
follows from $df=0$.

Notice that if we started with some other point $x_0$ (not the origin), then 
(\ref{Descent}) would change by $Q_{Lie}$ of something. For example, an
infinitesimal shift by $y$ would change (\ref{Descent}) by the $Q_{Lie}$-exact
expression:
\begin{equation}
{1\over 2} \iota_c^2 y^{\rho}\partial_{\rho} f(0) =
y^{\rho} c^{\mu} c^{\nu} \partial_{\rho} f_{\mu\nu}(0) =
-c^{\mu} c^{\nu} \partial_{\mu} f_{\nu\rho} y^{\rho} =
Q_{Lie} (f_{\nu\rho}c^{\nu}y^{\rho})
\end{equation}
(This is a manifestation of the general fact, that a Lie algebra acts trivially
in its cohomology.)

The $Q_{Lie}$-cohomology class of: 
\begin{equation}\label{Obstacle}
\iota_c^2 f(0) = f_{\mu\nu}(0) c^{\mu}c^{\nu}
\end{equation} 
is the obstacle for defining $a$ such that $f=da$ in a covariant way.

We have so far discussed only the action of shifts. The expression
(\ref{Obstacle}) as we defined it represents the cohomology class
of the algebra of translations ${\bf R}^{1+9}$. But we can also think
of it as a cocycle of the Poincare algebra. Indeed, $f_{\mu\nu}$
transforms covariantly under rotations and boosts and therefore
(\ref{Obstacle}) is closed under the $Q_{Lie}$ of the full Poincare algebra.

\vspace{7pt}

{\small \noindent
The $Q_{Lie}$ of the full Poincare algebra is the sum of $Q^{translations}_{Lie}$ of
translations ${\bf R}^{1+9}$ and $Q^{Lorentz}_{Lie}$ of rotations and boosts.
Expression (\ref{Obstacle}) is in the kernel of $Q^{translations}_{Lie}$ by
our construction, and more explicitly because $f$ is a closed form.
But it is also in the kernel of $Q^{Lorentz}_{Lie}$ because $f$
transforms covariantly under rotations and boosts.}

\vspace{7pt}

\noindent
Another question is whether or not (\ref{Obstacle}) is exact. One can see that
this is not exact as a cocycle of the full Poincare algebra, in the following
way. Let ${\bf P}$ stand for the Poincare algebra.  We have:
\begin{equation}
f_{\mu\nu}(0) c^{\mu}c^{\nu} \in H^2({\bf P},{\cal H}')
\end{equation}
Notice that the space of states of the gauge field contains a proper subspace closed
under the action of the Poincare algebra. (In other words, it is not an irreducible
representation.) This subspace consists of those gauge fields which have
a constant field strenght: $f_{\mu\nu}(x)=f_{\mu\nu}(0)$. Let us call this subspace
${\cal H}_{zero-modes}$:
\begin{equation}
{\cal H}_{zero-modes}\subset {\cal H}
\end{equation}
Therefore there is a projection 
\begin{equation}
{\cal H}'\to ({\cal H}_{zero-modes})'
\end{equation}
This projection naturally acts on the cocycles of $\bf P$ with values in ${\cal H}'$,
and therefore on the cohomology groups:
\begin{eqnarray}
&& {\cal H}'\otimes \Lambda^{\bullet} {\bf P} \to 
   ({\cal H}_{zero-modes})'\otimes \Lambda^{\bullet} {\bf P}
\nonumber \\
&& H^{\bullet}({\bf P},{\cal H}') \to 
   H^{\bullet}({\bf P},({\cal H}_{zero-modes})')
\end{eqnarray}
It is straightforward to see that the projection of (\ref{Obstacle}) to 
$({\cal H}_{zero-modes})'\otimes \Lambda^{\bullet} {\bf P}$ is automatically
a nonzero cohomology class. 

\vspace{7pt}
{\small\noindent
Indeed ${\cal H}'_{zero-modes}$ transforms as antisymmetric rank 2 tensor 
of the Lorentz algebra, and trivially under translations. 
Therefore the cohomology complex of the Poincare algebra with coefficients in
${\cal H}'_{zero-modes}$ is equivalent to the cohomology complex of the
Lorentz algebra with coefficients in 
$\Lambda^2{\bf R}^{1+9}\otimes \Lambda^{\bullet}{\bf R}^{1+9}$; the
projection of (\ref{Obstacle}) is in 
$H^0(Lorentz, (\Lambda^2{\bf R}^{1+9}\otimes \Lambda^2{\bf R}^{1+9})_{inv})$.
}
\vspace{7pt}

\noindent
This implies that (\ref{Obstacle}) represents a nontrivial
cohomology class in $H^2({\bf P},{\cal H}')$. This is what prevents us from
choosing the vertex covariant with respect to the Poincare algebra.

\subsubsection{The obstacle (\ref{Obstacle}) vanishes after we break
${\bf P}$ to ${\bf P}_{small}$}
Let us start with introducing some notations. 
For any vector $v^{\mu}$
we denote $\overline{v}$ the vector with the components:
\begin{equation}\label{BarNotation}
\overline{v}^{\mu} = \left\{
\begin{array}{c}
v^{\mu}\;\mbox{if } \mu\in\{0,1,\ldots,4\}\cr
-v^{\mu}\;\mbox{if  } \mu\in\{5,\ldots,9\}
\end{array}
\right.
\end{equation}
Also introduce:
\begin{equation}
2\Delta_S = \partial_{\mu}\overline{\partial}_{\mu} - \partial_{\mu}\partial_{\mu}
=2\sum_{i\in\{5,\ldots,9\}} \left({\partial\over\partial x^i}\right)^2
\end{equation}
What happens if we do not require the invariance under the full Poincare
algebra ${\bf P}$, but only under  the ${\bf P}_{small}$ of (\ref{PSmall})? 
Then we can restrict ourselves
to the subspace of ${\cal H}$ where $-\Delta_S$ is a fixed positive number.
On this subspace, it is possible to express $a_{\mu}$ in terms of $f_{\mu\nu}$
in a ${\cal P}_{small}$-covariant way. 
Let us choose the covariant gauge:
\begin{equation}
\partial^{\mu}a_{\mu} = 0
\end{equation}
and fix the residual gauge transformations with the additional ``axial'' gauge
gauge condition:
\begin{equation}\label{AxialGauge}
\overline{\partial}^{\mu} a_{\mu} = 0
\end{equation}
where $\overline{\partial}$ is introduced as in (\ref{BarNotation}). 
In the gauge (\ref{AxialGauge}) we can express the gauge field $a_{\mu}$ in terms of
the gauge field strength $f_{\mu\nu}=\partial_{\mu}a_{\nu}-\partial_{\nu}a_{\mu}$:
\begin{equation}\label{CovariantGauge}
a_{\mu}={1\over 2\Delta_S} \overline{\partial}_{\nu} f_{\mu\nu}
\end{equation}
Now we have two different expressions for the vector potential,
Eq. (\ref{VicinityGauge}) and Eq. (\ref{CovariantGauge}). The difference
between these two expressions is a gauge transformation. Let us use a ``diacritical'' 
mark to distinguish (\ref{CovariantGauge}) from (\ref{VicinityGauge}):
\begin{equation}
\acute{a}_{\mu} = {1\over 2\Delta_S} \overline{\partial}_{\nu} f_{\mu\nu}
\;\;\mbox{\it vs.}\;\;
a_{\mu}=\left(\iota_E {1\over {\cal L}_E} f \right)_{\mu}
\end{equation}
Note that $\acute{a}$ is ${\bf P}_{small}$-covariant:
\begin{equation}
Q_{Lie}\acute{a}=0
\end{equation}
while $a$ is not --- see Eq. (\ref{DescentDeRham}). 
Also, for every gauge field we can calculate $\acute{a}(0)$. This is
a functional of the gauge field, {\it i.e.} an element of ${\cal H}'$.
If we calculate its $Q_{Lie}$ as a cochain with values in ${\cal H}'$
we get:
\begin{equation}
Q_{Lie}(\;\acute{a}_{\mu}(0)\;) = - ({\cal L}_c \acute{a}_{\mu}) (0)
\end{equation}

\vspace{7pt}
\noindent
Now let us return to Eqs. (\ref{DescentDeRham}), (\ref{EqualsConstant})
and (\ref{Descent}). On the subspace
$-\Delta_S=\mbox{const}>0$ the cohomology class of (\ref{Descent}) trivializes:
\begin{equation}
c^{\mu}c^{\nu} f_{\mu\nu}(0) 
= Q_{Lie} 
\left( \iota_c \acute{a}(0) \right)
\end{equation}
This is analogous to Eq. (\ref{H3Zero}) of Section \ref{sec:Descent}.
Therefore the same arguments as we presented in Section \ref{sec:Descent}
should imply that 
\begin{equation}
{1\over {\cal L}_E({\cal L}_E+1)} \iota_E\iota_c f(x) - 
\iota_c \acute{a}(0) = Q_{Lie}(\mbox{something})
\end{equation}
Indeed we have:
\begin{eqnarray}
&& {1\over {\cal L}_E({\cal L}_E+1)} \iota_E\iota_c f(x) 
-\iota_c \acute{a}(0)
=
\\
& = &
Q_{Lie} \left( - \iota_E\left(
\acute{a}(0)+{1\over {\cal L}_E}(\acute{a}-\acute{a}(0))
\right)\right)
\nonumber
\end{eqnarray}
Also notice that:
\begin{equation}
d\left(\iota_E\left(
\acute{a}(0)+{1\over {\cal L}_E}(\acute{a}-\acute{a}(0))
\right)\right) = \acute{a}-a
\end{equation}
This means that the correction of the vector potential:
\begin{equation}
a\to \acute{a}
\end{equation}
is completely analogous to the correction of the vertex operator
described in Section \ref{sec:Descent}. It turns the non-covariant
expression $a_{\mu}dx^{\mu}$ into the covariant expression
$\acute{a}_{\mu} dx^{\mu}$.

\subsubsection{Action of the supersymmetry}
Let us now study the action of the supersymmetry generators. Let us consider
the part of the $Q_{Lie}$ involving the supersymmetry generators.
We have:
\begin{eqnarray}
Q_{Lie} &  = & Q_{Lie}^{(x,\theta)} + Q_{Lie}^{\cal H} \\
Q_{Lie}^{(x,\theta)} &  = & \xi^{\alpha}{\partial\over\partial\theta^{\alpha}}
-\xi^{\alpha}\Gamma^{\mu}_{\alpha\beta}\theta^{\beta}{\partial\over\partial x^{\mu}}
\\
Q_{Lie}^{\cal H} & = & \xi^{\alpha}t_{\alpha}
\\[5pt]
Q_{BRST} & = & \lambda^{\alpha}{\partial\over \partial\theta^{\alpha}}
+ \lambda^{\alpha}\Gamma^{\mu}_{\alpha\beta}\theta^{\beta} {\partial\over \partial x^{\mu}}
\end{eqnarray}
Here $t_{\alpha}$ is the supersymmetry transformation acting on the space of states
${\cal H}$ and therefore (after fixing the gauge!) on $a_{\mu}$ and $\psi$. 
We write only the part of $Q_{Lie}$ corresponding to  the  super-translations;
$\xi^{\alpha}$ are the bosonic ghosts corresponding to  the super-translations.
To make $t_{\alpha}$ act on $a$ and $\psi$ we have to choose the gauge.

With this notation, let us  first of all present $Q_{Lie}^{(x,\theta)}$ acting
on the vertex operator (\ref{VertexOperator}) in the following
form:
\begin{eqnarray}
 \varepsilon'Q_{Lie}^{(x,\theta)} \varepsilon V & = &
(\varepsilon\lambda\Gamma^{\mu}\varepsilon'\xi)a_{\mu} -
(\varepsilon'\xi \Gamma^{\nu} \theta)(\varepsilon\lambda\Gamma^{\mu}\theta) \partial_{\nu} a_{\mu} -
\nonumber
\\
&& 
-{1\over 2}  (\varepsilon'\xi\Gamma^{\mu\nu\rho}\theta)(\varepsilon\lambda \Gamma_{\rho} \theta) 
\partial_{[\mu}a_{\nu]} 
-{1\over 4}  (\theta\Gamma^{\mu\nu\rho}\theta)(\varepsilon\lambda\Gamma_{\rho} \varepsilon'\xi) 
\partial_{[\mu}a_{\nu]} +
\nonumber
\\
&& +
(\varepsilon\lambda\Gamma^{\mu}\varepsilon'\xi)(\psi\Gamma_{\mu}\theta) + 
(\varepsilon\lambda\Gamma^{\mu}\theta)(\psi\Gamma_{\mu}\varepsilon'\xi) +\ldots =
\nonumber
\\[10pt]
 & = & - {3\over 2}(\varepsilon\lambda\Gamma^{\mu}\theta)
\left( (\varepsilon'\xi\Gamma_{\mu}\psi) 
+ {1\over 2\Delta_S}\partial_{\mu} (\varepsilon'\xi\overline{\Gamma}^{\rho}\partial_{\rho}\psi)\right) -
\nonumber
\\
&& -{2\over 3} (\varepsilon\lambda\Gamma^{\rho}\theta)(\varepsilon'\xi\Gamma^{\mu\nu}\Gamma_{\rho}\theta)
\partial_{[\mu}a_{\nu]} + \ldots
\\
&& + \varepsilon Q_{BRST} 
\left( {1\over 2} (\theta\Gamma^{\mu}\varepsilon'\xi)(\psi\Gamma_{\mu}\theta)
+ {3\over 2}{1\over 2\Delta_S} (\varepsilon'\xi\overline{\Gamma}^{\mu}\partial_{\mu}\psi)
+ \right.
\nonumber
\\
&& \phantom{+ Q_{BRST}\;\;} 
\left.
+(\theta\Gamma^{\mu}\varepsilon'\xi) a_{\mu}
-{1\over 12} (\theta\Gamma^{\mu\nu\rho}\theta) (\theta\Gamma_{\rho}\varepsilon'\xi)
\partial_{[\mu}a_{\nu]}
+\ldots \right)
\nonumber
\end{eqnarray}
This implies that 
\begin{eqnarray}
\varepsilon'Q_{Lie}^{\cal H} a_{\mu} & = &
-{3\over 2} \left((\varepsilon'\xi\Gamma_{\mu}\psi)
+\partial_{\mu} {1\over 2\Delta_S}(\varepsilon'\xi\overline{\Gamma}^{\rho}\partial_{\rho}\psi)\right)
\\
\varepsilon'Q_{Lie}^{\cal H} \psi & = & - {2\over 3}  \varepsilon'\xi \Gamma^{\mu\nu}\partial_{[\mu}a_{\nu]}
\end{eqnarray}
Therefore we have indeed:
\begin{equation}
\varepsilon'Q_{Lie} \varepsilon V = 
\varepsilon Q_{BRST}\left( (\theta\Gamma^{\mu}\varepsilon'\xi) a_{\mu}  
+{3\over 2} {1\over 2\Delta_S}
(\varepsilon'\xi\overline{\Gamma}^{\mu}\partial_{\mu}\psi) + \ldots\right)
\end{equation}
On the right hand side $Q_{BRST}$ is taken of the  expression  which is
$Q_{Lie}$-exact:
\begin{eqnarray}
&& (\theta\Gamma^{\mu}\varepsilon'\xi) a_{\mu}  
+ {3\over 2} {1\over 2\Delta_S}(\varepsilon'\xi\overline{\Gamma}^{\mu}\partial_{\mu}\psi) + \ldots
=
\nonumber
\\
&& =\varepsilon'Q_{Lie}\left(
{3\over 2} {1\over 2\Delta_S}(\theta\overline{\Gamma}^{\mu}\partial_{\mu}\psi) 
-{1\over 2}{1\over 2\Delta_S}
(\theta\Gamma^{\mu\rho\sigma}\theta) \overline{\partial}_{\mu} \partial_{\rho} a_{\sigma}
+\ldots \right)
\end{eqnarray}
Then we have:
\begin{eqnarray}
&&  Q_{BRST}  \left(
{3\over 2} {1\over 2\Delta_S}(\theta\overline{\Gamma}^{\mu}\partial_{\mu}\psi) 
-{1\over 2}{1\over 2\Delta_S}
(\theta\Gamma^{\mu\rho\sigma}\theta) \overline{\partial}_{\mu} \partial_{\rho} a_{\sigma}
+\ldots \right) = 
\\
&& \;\;\; = {1\over 2\Delta_S}\left({3\over 2} (\lambda\overline{\Gamma}^{\mu}\partial_{\mu}\psi)
+ {3\over 2} (\theta\Gamma^{\nu}\lambda) (\theta\overline{\Gamma}^{\mu}\partial_{\mu}\partial_{\nu}\psi) 
- (\lambda\Gamma^{\mu\rho\sigma}\theta) \overline{\partial}_{\mu} \partial_{\rho} a_{\sigma}
+\ldots \right)
\nonumber
\end{eqnarray}
This means that the following vertex operator:
\begin{equation}
\tilde{V}=
-{3\over 2} {1\over 2\Delta_S} 
(\lambda\overline{\Gamma}^{\mu}\partial_{\mu}\psi) + 
(\lambda\Gamma^{\mu}\theta)a_{\mu} + 
{1\over 2\Delta_S}
(\lambda\Gamma^{\mu\rho\sigma}\theta) \partial_{\rho} \overline{\partial}_{\mu} a_{\sigma}
+\ldots
\end{equation}
transforms covariantly under the odd shifts.
This can be also understood as follows:
\begin{equation}
\tilde{V} = -{3\over 2} {1\over 2\Delta_S} 
(\lambda\overline{\Gamma}^{\mu}\partial_{\mu}\psi) +
(\lambda\Gamma^{\rho}\Gamma^{\mu\nu}\theta)
{1\over 2\Delta_S}
\overline{\partial}_{\rho} \partial_{[\mu}a_{\nu]} 
+\ldots
\end{equation}
Now we recognize what it is:
\begin{equation}\label{MaxwellVertexViaW}
\tilde{V}(x,\theta,\lambda) = 
{1\over 2\Delta_S} 
\lambda^{\alpha} \overline{\Gamma}^{\mu}_{\alpha\beta} 
\partial_{\mu} W^{\beta}(x,\theta)
\end{equation}
where $W^{\alpha}(x,\theta)$ is the superfield\footnote{I want to thank
Yuri Aisaka for duscussions about $W^{\alpha}$} related to the Maxwell superfield
$A_{\alpha}(x,\theta)$ by the chain of transformations:
\begin{eqnarray}
{\cal T}_{(\alpha}A_{\beta)} & = & {1\over 2} \Gamma_{\alpha\beta}^{\mu} A_{\mu} \\ 
{\cal T}_{\alpha} A_{\mu} - {\cal T}_{\mu} A_{\alpha} & = &
g_{\mu\nu}\Gamma^{\nu}_{\alpha\beta} W^{\beta}
\end{eqnarray}
See  \cite{Mafra:2009wq} for a recent discussion of $W^{\alpha}$.

\subsection{Supergravity field.}
In flat space the supergravity
fields split into the product of the left and the right component; 
the left and right components are essentially free Maxwell fields. 
The bispinor field is defined as follows:
\begin{equation}
P^{\alpha\dot{\beta}} = W_L^{\alpha} W_R^{\dot{\beta}}
\end{equation}
where $W$ is the field strength superfield of the free Maxwell theory; the
$\theta=0$ component of $W^{\alpha}$ is the  gaugino $\psi^{\alpha}$.

This bispinor field is a linear combination of the RR field strengths 
contracted with the gamma-matrices \cite{Berkovits:2001ue}:
\begin{eqnarray}
P^{\alpha\dot{\beta}} & = & F_{\tang{l}\tang{m}\tang{n}\tang{p}\tang{q}}
e^{\tang{l}}_a e^{\tang{m}}_b e^{\tang{n}}_c e^{\tang{p}}_d e^{\tang{q}}_e
(\Gamma^{abcde})^{\alpha\dot{\beta}} + 
\nonumber
\\
& + & a_3 F_{\tang{l}\tang{m}\tang{n}}
e^{\tang{l}}_a e^{\tang{m}}_b e^{\tang{n}}_c
(\Gamma^{abc})^{\alpha\dot{\beta}}
+
a_1 F_{\tang{l}} e^{\tang{l}}_a (\Gamma^{a})^{\alpha\dot{\beta}}
\label{BispinorRR}
\end{eqnarray}
where $a_3$ and $a_1$ are some numeric coefficients which we do not need.
The supersymmetry variations of $P$ is given by this equation:
\begin{eqnarray}
t^3_{\alpha}P^{\beta\dot{\beta}} = 
\delta_{\alpha}^{\beta}C^{\dot{\beta}} + (\Gamma_{mn})_{\alpha}^{\beta} C^{\dot{\beta}mn}
\label{BispinorSUSY_left}
\\
t^1_{\dot{\alpha}}P^{\beta\dot{\beta}} = 
\delta_{\dot{\alpha}}^{\dot{\beta}}C^{\beta} + 
(\Gamma_{mn})_{\dot{\alpha}}^{\dot{\beta}}C^{\beta mn}
\label{BispinorSUSY_right}
\end{eqnarray}
where $C^{\beta}$ is a combination of the left dilatino $\chi$, and the left 
gravitino field
strength $\partial_{[m}\psi_{n]}$, and $C^{\dot{\beta}}$ is a combination of 
the corresponding right fields
$\tilde{\chi}$ and $\partial_{[m}\tilde{\psi}_{n]}$.

The ${\bf SP}_{small}$-covariant vertex in flat space is the product of two
expressions of the form (\ref{MaxwellVertexViaW}):
\begin{equation}\label{VertexWithDerivatives}
\tilde{V} = (\lambda \overline{\Gamma}^{\mu} \;
\Delta_S^{-2} \partial_{\mu}\partial_{\nu}P \; \overline{\Gamma}^{\nu} \tilde{\lambda})
\end{equation}
This is BRST equivalent to\footnote{I want to thank N.~Berkovits for suggesting this
simplified form, and many other useful comments}:
\begin{equation}\label{VertexWithoutDerivatives}
\tilde{V}' = (\lambda \overline{\Gamma}^{\mu} \Delta_S^{-1} P \Gamma_{\mu} \tilde{\lambda})
\end{equation}

\subsection{Relation between the covariant vertex in $AdS_5\times S^5$
and the flat space expressions (\ref{VertexWithDerivatives}), 
(\ref{VertexWithoutDerivatives}).}
The construction of Section \ref{sec:CohomologicalArgument} implies that the 
$PSU(2,2|4)$-covariant vertex
operator exists in $AdS_5\times S^5$. 
This construction gives (\ref{VertexWithDerivatives}) when applied
in the flat space limit. (We have demonstrated this for the Maxwell field,
but the free supergravity vertex in flat space is just the product
of ``left'' and ``right'' Maxwell vertices.)
Therefore both 
(\ref{VertexWithDerivatives}) and  the BRST-equivalent (\ref{VertexWithoutDerivatives})
should be the flat space limits of some covariant vertices in $AdS_5\times S^5$.

Notice that (\ref{VertexWithoutDerivatives}) reduces to 
(\ref{ZeroModeVertex}) in the zero momentum limit, except for the
overall normalization factor $1\over \Delta_S$ which becomes singular
on the zero mode. 

However we were not able to write explicit expressions in terms
of the supergravity fields in AdS space which would explicitly generalize 
the flat space formulas 
(\ref{VertexWithDerivatives}) or (\ref{VertexWithoutDerivatives}).

\section{Covariant vertex and the endpoint of the  Wilson line}
\label{sec:RelationToEndpoint}

\subsection{The BRST complex of the endpoint}
Consider the Wilson line operator corresponding to a
semi-infinite contour going from infinity to
some point $B$ on the string worldsheet:

\vspace{10pt}

\hspace{0.8in}\includegraphics[width=2.4in]{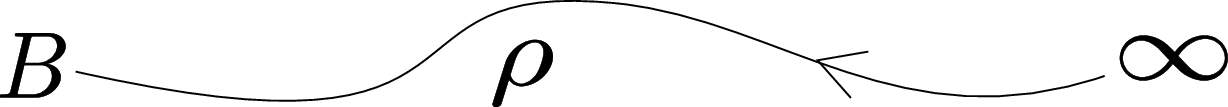}

\vspace{10pt}

\noindent
in some representation $\rho$ of ${\bf psu}(2,2|4)$.
Consider the action of $Q_{BRST}$ on this operator. If we neglect
the contribution of the boundary terms at infinity, then the
BRST variation is\cite{Vallilo:2003nx,Berkovits:2004jw,Berkovits:2004xu}:
\begin{equation}\label{BoundaryTermsSemiInfinite}
\varepsilon Q_{BRST} T[C_{\infty}^B](z)  = 
\left( {1\over z} \varepsilon\lambda^{\alpha}_3(B)\rho(t^3_{\alpha}) 
+ z \varepsilon\lambda_1^{\dot{\alpha}}(B)\rho(t^1_{\dot{\alpha}}) \right) 
T[C_{\infty}^B](z) 
\end{equation}
(See 
 Section 2.2 of \cite{Mikhailov:2007mr} and  Section 7 of \cite{Mikhailov:2007eg}
 for a discussion  of this formula.)

Let us fix some vector $\psi$ in the representation $\rho$ ``at infinity'';
then this expression:
\begin{equation}\label{VectorFromInfinity}
T[C_{\infty}^B](z)\psi
\end{equation}
is a vector in the representation space of $\rho$.
Pick a vector $v$ in the dual space, and evaluate it on (\ref{VectorFromInfinity}):
\begin{equation}
v\left(T[C_{\infty}^B](z)\psi\right) \;\in\; {\bf C}
\end{equation}
This gives a number. Consider vectors $v$ depending on the pure
spinors $\lambda_3,\lambda_1$ and the spectral parameter $z$.
Then Eq. (\ref{BoundaryTermsSemiInfinite}) can be regarded as
defining the action of $Q_{BRST}$ on $v$:
\begin{equation}\label{BRSTComplexEndpoint}
Q_{endpoint}v=
\left({1\over z}\lambda_3^{\alpha}\rho(t^3_{\alpha})+
z \lambda_1^{\dot{\alpha}}\rho(t^1_{\dot{\alpha}})\right) v
\end{equation}
This defines the BRST complex of the endpoint. 
The $n$-cochains of this complex are elements
\[
v\in 
\mbox{Hom}_{{\bf g}_{\bar{0}}}
\left(
\left[
\begin{array}{l}
\mbox{linear space of the}\cr
\mbox{representation $\rho$ in which}\cr
\mbox{we evaluate Wilson line}
\end{array}
\right]
\; , \; 
{\cal P}^n
\right)
\]
where ${\cal P}^n$ is defined in Section \ref{sec:PureSpinors},
and the differential is (\ref{BRSTComplexEndpoint}).
The ``plugs'' which we introduced in Section \ref{sec:ThePlugs}
are the cohomologies of this complex.
Unfortunately
we do not know a general classification of the cohomologies
of this complex for a general representation $\rho$.

\subsection{The BRST complex of the Wilson line endpoint
is isomorphic to the BRST complex of covariant vertices}

We will now consider the special case where $\rho$ is the
representation of $psu(2,2|4)$ on the space of states ${\cal H}$
of the BPS multiplet.
In this case
we will relate  the BRST complex of the endpoint (\ref{BRSTComplexEndpoint})
to the BRST complex of covariant supergravity
vertices. 

\subsubsection{Frobenius reciprocity}
Let us remember the general structure of the covariant
vertex from Section \ref{sec:CovariantUniversal}:
\[
{\cal V} \in \mbox{Hom}_{\bf g}({\cal H}, 
\mbox{Coind}_{{\bf g}_0}^{\bf g}{\cal P}^n)
=
\mbox{Hom}_{\bf g} ({\cal H}, 
                    \mbox{Hom}_{{\bf g}_{\bar{0}}}({\cal U}{\bf g},
                                                   {\cal P}^n))
\]
Here $n=2$ for the supergravity vertex, but we want
to consider the whole BRST complex so we keep $n$.
Let us evaluate ${\cal V}$ on the unit of the group:
\begin{equation}
\mbox{Hom}_{\bf g} ({\cal H}, 
  \underbrace{\mbox{Hom}_{{\bf g}_{\bar{0}}}({\cal U}{\bf g},
  {\cal P}^n)}_{\mbox{\small evaluate on }
  \displaystyle{\bf 1}\in {\cal U}{\bf g}})
\end{equation}
This defines a correspondence between covariant vertices ${\cal V}$
and vectors in $\mbox{Hom}_{{\bf g}_{\bar{0}}} ({\cal H}, {\cal P}^n)$:
\begin{equation}\label{EvaluationAtZero}
\mbox{Hom}_{\bf g}({\cal H}, 
\mbox{Coind}_{{\bf g}_0}^{\bf g}{\cal P}^n)
\owns {\cal V} \;\;
\stackrel{\mbox{evaluate on }{\displaystyle\bf 1}}{\longrightarrow}
\;\; v \in \mbox{Hom}_{{\bf g}_{\bar{0}}} ({\cal H}, {\cal P}^n)
\end{equation}
Notice that ${\cal V}$ is a function of $x,\theta$ while $v$
is essentially its value at $x=\theta=0$. Nevertheless, the correspondence
(\ref{EvaluationAtZero}) is a one-to-one correspondence between
the elements of 
$\mbox{Hom}_{\bf g}({\cal H},
                    \mbox{Coind}_{{\bf g}_{\bar{0}}}^{\bf g}{\cal P}^n)$ 
and the elements
of $\mbox{Hom}_{{\bf g}_{\bar{0}}} ({\cal H}, {\cal P}^n)$. 
Indeed, the symmetry under
$\bf g$:
\[
\mbox{Hom}_{\underbrace{\bf g}_{\mbox{\small this }{\displaystyle \bf g}}}
({\cal H},\mbox{Coind}_{{\bf g}_0}^{\bf g}{\cal P}^n)
\]
allows to relate ${\cal V}(\Psi)(g)$ to ${\cal V}(g^{-1}\Psi)({\bf 1})$,
see Eq. (\ref{ActionOfGOnCovariant}). In other words, if we know
the value of the covariant vertex at the point $x=\theta=0$ 
{\em for all states } $\Psi$, then because of the global symmetry 
we know the covariant vertex
everywhere (for arbitrary $x$ and $\theta$). 
This construction is an example of the {\em Frobenius reciprocity}:
\begin{center}
\fbox{$
\mbox{Hom}_{\bf g}({\cal H}, 
\mbox{Coind}_{{\bf g}_0}^{\bf g}L)
\simeq
\mbox{Hom}_{{\bf g}_{\bar{0}}} 
({\cal H}|_{{\bf g}_{\bar{0}}}, L)
$}
\end{center}
which is true for any representation $L$ of ${\bf g}_{\bar{0}}$;
in our case $L={\cal P}^n$.

To summarize, given the covariant vertex ${\cal V}$, we define
$v\in \mbox{Hom}_{{\bf g}_{\bar{0}}}({\cal H}, {\cal P}^n)$
by saying that the value of $v$ on $\Psi\in {\cal H}$ is:
\begin{equation}
v(\Psi) =  {\cal V}(\Psi)({\bf 1}) 
\label{VectorFromCovariantVertex}
\end{equation}

\subsubsection{The action of $Q_{BRST}$}
\label{sec:DefinitionOfAlgebraicComplex}
Note that both the left hand side and the right hand side of 
(\ref{VectorFromCovariantVertex}) are elements of ${\cal P}^n$
{\it i.e.} polynomials of $\lambda_3$ and $\lambda_1$. 
We can evaluate them on $\lambda$:
\begin{equation}
v(\Psi)(\lambda_3,\lambda_1) =  
{\cal V}(\Psi)({\bf 1})(\lambda_3,\lambda_1)
\end{equation}
This is a quadratic polynomial in $\lambda_3$ and $\lambda_1$.
Eqs. (\ref{ActionOfGOnCovariant}) and (\ref{QonUniversal})  imply 
that the action of $Q_{BRST}$ on the covariant
vertex corresponds to the following action on $v$:
\begin{eqnarray}
&&  (Q_{BRST} v) (\Psi) (\lambda_3,\lambda_1)= 
- v (\lambda_3\Psi  + \lambda_1\Psi)(\lambda_3,\lambda_1)
\label{BRSTonCovariantVertex}
\end{eqnarray}
This formula for $Q_{BRST}$ can be interpreted in the following way.
The space $\mbox{Hom}_{\bf C}({\cal H},{\cal P}^2)$ is obviously a representation
of ${\bf g}=psu(2,2|4)$, just because ${\cal H}$ is by definition a representation
of ${\bf g}$. (The ${\cal P}^2$ part just
``goes along for the ride''.) Let us denote this representation $\rho$
(the action of $x\in{\bf g}$ on $v$ is $\rho(x)v$). Then (\ref{BRSTonCovariantVertex})
implies that the action of ${\cal Q}_{BRST}$ on 
$\mbox{Hom}_{\bf g}({\cal H},  \mbox{Coind}_{{\bf g}_0}^{\bf g}{\cal P}^2)$ corresponds
to the action of the nilpotent operator $Q_{endpoint}$ on $v$ defined by this formula:
\begin{equation}\label{ZEqualsOne}
Q_{BRST}v(\lambda_3,\lambda_1)=(\lambda_3^{\alpha}\rho(t^3_{\alpha})+
\lambda_1^{\dot{\alpha}}\rho(t^1_{\dot{\alpha}}))v(\lambda_3,\lambda_1)
\end{equation}
This is identical to $Q_{endpoint}$ of (\ref{BRSTComplexEndpoint}) at
$z=1$.  
We conclude that:
\begin{quote}
 $v$ represents a cohomology class $H^2(Q_{endpoint})$ of the following complex: 
\\
\begin{equation}
\ldots \longrightarrow 
{\cal H}'\otimes_{{\bf g}_{\bar{0}}} {\cal P}^n 
\stackrel{Q_{endpoint}}{\longrightarrow}
{\cal H}'\otimes_{{\bf g}_{\bar{0}}} {\cal P}^{n+1} 
\longrightarrow \ldots
\end{equation}
\end{quote}
In other words, the  BRST complex on covariant massless vertices (independent
of derivatives) is equivalent to the endpoint complex 
$\mbox{Hom}_{{\bf g}_0}({\cal H},{\cal P}^{\bullet})$.

\subsubsection{Including the spectral parameter $z$ corresponds
to rescaling the pure spinors.}
We have demonstrated that the complex (\ref{ZEqualsOne}) is equivalent
to (\ref{BRSTComplexEndpoint}) at $z=1$. But in fact 
(\ref{BRSTComplexEndpoint}) at $z=1$ is equivalent to 
(\ref{BRSTComplexEndpoint}) at $z\neq 1$ by rescaling of $\lambda_3$
and $\lambda_1$. In other words, the map
\begin{eqnarray}
&& v \mapsto v' 
\nonumber
\\
&& v'(\lambda_3,\lambda_1) = v(z^{-1}\lambda_3,z\lambda_1)
\end{eqnarray}
is the equivalence of the complex (\ref{BRSTComplexEndpoint})
at $z=1$ and the same complex at $z\neq 1$.

\subsection{Endpoint BRST complex and the Lie algebra cohomology}
There is a relation between the endpoint cohomology and the 
cohomology of the positive-frequency part of the loop algebra of
$psl(4|4)$.

Consider the algebra formed by the positive frequency ${\bf Z}_4$-twisted loops with
values in $psl(4|4)$. We will denote is  $L_+{\bf g}$. 
The cohomology complex is generated by the ghosts
$c_{-k}^a$, where $k\in \{1,2,3,\ldots\}$ and $a$ the enumerates the adjoint
representation of $psl(4|4)$. We have $c_{-1}^{\alpha}$, $c_{-2}^m$, 
$c_{-3}^{\dot{\alpha}}$, $c_{-4}^{[mn]}$, $c_{-5}^{\alpha}$, {\it etc.}
The ``energy'' operator $L_0$ counts the lower indices, for example:
\[
L_0 c_{-3}^{\dot{\alpha}} = -3 c_{-3}^{\dot{\alpha}}\;,\;\;
L_0 c_{-1}^{\alpha} c_{-4}^{[mn]} = -5 c_{-1}^{\alpha} c_{-4}^{[mn]}
\] 
Notice that $L_0$ is a symmetry of the cohomology complex. Another symmetry
is the $c$-ghost number (the number of letters $c$). 
Let $H^p_q(L_+{\bf g},{\bf C})$ denote the cohomology group with $L_0 = q$
and ghost number $p$.
The first cohomology group $H^1(L_+{\bf g},{\bf C})$ is generated
by $c^{\alpha}_{-1}$ (and therefore has $L_0=-1$). Some of other nontrivial 
cohomology groups are:
\begin{eqnarray}
H^2_{-2}(L_+{\bf g},{\bf C}): & \;\;\; & X_{\alpha\beta} c_{-1}^{\alpha} c_{-1}^{\alpha},\;\;\;\;
{f_m}^{\alpha\beta} X_{\alpha\beta} = 0
\label{H2OfL}
\\
H^3_{-3}(L_+{\bf g},{\bf C}): & \;\;\; & X_{\alpha\beta\gamma} 
c_{-1}^{\alpha} c_{-1}^{\beta} c_{-1}^{\gamma},\;\;\;\;
{f_m}^{\alpha\beta} X_{\alpha\beta\gamma} = 0
\label{H3OfL}
\end{eqnarray}
Generally speaking, $H^k_{-k}$ is generated by the expressions of the form:
\begin{equation}
\label{WrongFormula}
H^k_{-k}(L_+{\bf g},{\bf C}):  \;\;\;  X_{\alpha_1\cdots\alpha_k} 
c_{-1}^{\alpha_1} \cdots c_{-1}^{\alpha_k} ,\;\;\;\;
{f_m}^{\alpha_1\alpha_2} X_{\alpha_1\alpha_2\cdots\alpha_k} = 0
\end{equation}
But $H^k_{-k}$ is not all of the cohomology, 
for example there is nontrivial\footnote{this can be
seen from the comparison of the character
of the $H^{\bullet}(L_+{\bf g},{\bf C})$ with the
character of \cite{Berkovits:2005hy}} $H^2_{-4}$. 
The pure spinor cohomology should be identified with
the part of the $L_+{\bf g}$ cohomology with ``energy'' equal
to minus the ghost number, {\it i.e.} $H^k_{-k}$.

\section{Application: vertex operators depending on the spectral parameter}
\label{sec:Applications}

\subsection{How to introduce the spectral parameter into the vertex operator}
In flat infinite space massless vertex operators have the form:
\begin{equation}\label{VertexInFlatSpace}
V_k(x,\theta) = p(\lambda,\theta) e^{ikX}
\end{equation}
where $p$ is some polynomial of $\lambda$ and $\theta$.
We can write
\begin{equation}\label{LeftRightDecomposition}
X(\tau^+,\tau^-)=
X_L(\tau^+) + X_R(\tau^-)
\end{equation}
where
\begin{equation}
X_L=\int_{\infty}^{(\tau^+,\tau^-)} d\tau^+ \partial_+ X 
\;\mbox{and }\;
X_R=\int_{\infty}^{(\tau^+,\tau^-)} d\tau^- \partial_- X
\end{equation}
Therefore there is a generalization of (\ref{VertexInFlatSpace}):
\begin{equation}\label{FlatSpaceGeneralizedVertex}
V_{k_L,k_R}(x,\theta) = p(\lambda,\theta) e^{ik_LX_L + ik_RX_R}
\end{equation}
This generalization is only well defined in flat space on those worldsheets
which do not have handles, or in toroidal compactifications with appropriate
integrality conditions on $k_L$ and $k_R$. We can formally
consider (\ref{FlatSpaceGeneralizedVertex}), for example on an infinite
worldsheet without handles, if we neglect boundary effects.

We will now argue that there is a partial analogue of
(\ref{FlatSpaceGeneralizedVertex}) in $AdS_5\times S^5$. 

Given a state $\Psi\in {\cal H}$ we can prepare a nonlocal $z$-dependent
covariant vertex operator, in the following way. Consider the transfer matrix
$T_{\infty}^{(\tau^+,\tau^-)}(z)$ from infinity to the point $(\tau^+,\tau^-)$
on the worldsheet. Let us fix a vector $\Psi^{(\infty)}\in {\cal H}$, and consider:
\begin{equation}\label{ZDependentVertex}
V_{\Psi_{\infty}}(\tau^+,\tau^-|z) = 
v\left( T^{(\tau^+,\tau^-)}_{(\infty)}(z)\Psi^{(\infty)} \right) 
(\;z^{-1}\lambda_3(\tau^+,\tau^-) \;,\; z\lambda_1(\tau^+,\tau^-)\;)
\end{equation}
So defined $V_{\Psi_{\infty}}(\tau^+,\tau^-|z)$ is analogous to:
\begin{equation}
e^{ikz^{-1}X_L + ikzX_R}
\end{equation}
In particular for $z=1$ we get 
$T^{(\tau^+,\tau^-)}_{(\infty)}(z)=g(\tau^+,\tau^-)g(\infty)^{-1}$ 
and therefore:
\begin{equation}
V_{\Psi_{\infty}}(\tau^+,\tau^-|1) = 
v\left(g(\tau^+,\tau^-)g(\infty)^{-1}\Psi^{(\infty)}\right) 
(\;\lambda(\tau^+,\tau^-)\;)
\end{equation}
This formula gives us back the covariant vertex for the state $\Psi$ if we
identify:
\begin{equation}
g(\infty)^{-1}\Psi^{(\infty)} = \Psi
\end{equation}

\subsection{Is there a 2-point vertex operator?}
\label{sec:TwoPoint}
For any $u\in {\cal H}'$ we define $u^{\dagger}$ as a non-normalizable
vector in the space of states, characterized by the formula:
\begin{equation}
u(\Psi) = (u^{\dagger},\Psi) \;\; \mbox{for any}\;\; \Psi\in {\cal H}
\end{equation}
where $(,)$ is the hermitean scalar product in ${\cal H}$. Note that
$u^{\dagger}$ strictly speaking does not belong to ${\cal H}$ because it is
not normalizable.
For example, one dimensional quantum mechanics has ${\cal H}=L^2({\bf R})$
--- the space of square integrable functions of one variable,
with the norm $|| f ||^2 = \int dx |f(x)|^2 $. The dual space ${\cal H}'$
is the space of generalized functions; if $u\in {\cal H}'$ is defined
by the formula $u(f)=f(0)$ then $u^{\dagger}$
is a delta-function $\delta(x)$.

Using these notations we can define the two-point vertex operator:

\begin{center}
\includegraphics[width=2.0in]{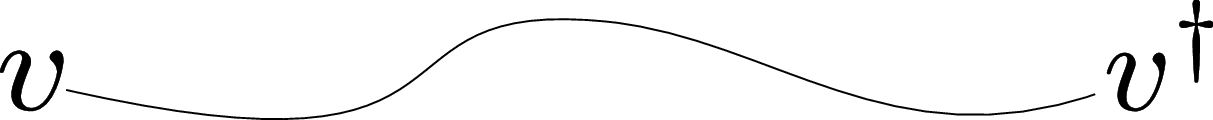}
\end{center}

\begin{equation}\label{TwoPointVertex}
V^{2pt}((\tau_1^+,\tau_1^-),(\tau_2^+,\tau_2^-)) = 
v\left(\; T^{\tau_1}_{\tau_2} v^{\dagger}(\lambda(\tau_2^+,\tau_2^-))\; \right) \;
(\;\lambda(\tau_1^+,\tau_1^-)\;)
\end{equation}
However, we conjecture that this 2-point vertex operator is in fact
BRST exact. Indeed, although we have not checked it explicitly,
it should be true that the derivative of 
$V^{2pt}((\tau_1^+,\tau_1^-),(\tau_2^+,\tau_2^-))$ with respect
to $\tau_1$ is $Q_{BRST}$-exact. Therefore, up to BRST-exact terms
this vertex is independent of $\tau_1$ and $\tau_2$. On the other
hand, when $\tau_1\to \tau_2$ we get a local vertex operator of
the ghost number 4. There is no $psu(2,2|4)$-invariant
cohomology at the ghost number 4.
This implies that (\ref{TwoPointVertex}) is BRST exact.

\section{Conclusions}
\label{sec:Conclusions}
In this paper we introduced a family of $z$-dependent 
vertex operators (\ref{ZDependentVertex}) parametrized by a choice
of the BPS representation of $psu(2,2|4)$. Schematically, these
vertex operators have a form:
\begin{equation}
V_{\Psi_{\infty}}(\tau^+,\tau^-|z)
=
\left\langle \mbox{plug}(\tau^+,\tau^-|z) \left| 
P\exp\left( -\int_{\infty}^{(\tau^+,\tau^-)} J[z] \right) 
\right|\Psi^{\infty}\right\rangle
\end{equation}
This expression is strictly speaking not BRST invariant, because of the
boundary term at infinity. Indeed we have put $\Psi^{\infty}$ an arbitrary
vector from ${\cal H}$, and this is generally speaking not a valid plug.
We assume that we can neglect this boundary term 
because it is at infinity\footnote{an attempt to bring the second endpoint
from infinity is described in Section \ref{sec:TwoPoint}}.
We can consider $V_{\Psi_{\infty}}(\tau^+,\tau^-|z)$ locally near the
point $(\tau^+,\tau^-)$. Notice that $\langle \mbox{plug}(\tau^+,\tau^-)|$
is a $\lambda$-dependent vector in the dual space to ${\cal H}$. 
(In fact $\langle \mbox{plug}(\tau^+,\tau^-)|$ depends on $\tau^+$ and
$\tau^-$ through $\lambda(\tau^+,\tau^-)$.)
We can
also think of $\langle \mbox{plug}(\tau^+,\tau^-)|$ as an element of ${\cal H}$, 
but then we have to remember
that it is not normalizable; it is a $\delta$-function type of state, rather
than a proper wave packet. 
Note that for a fixed $\lambda$, our $\langle \mbox{plug} |$ is a {\em fixed}
vector in ${\cal H}'$. In other words, for every BPS representation ${\cal H}$
we have a map, which takes a pair of pure spinors and transforms
them into a vector in the space of BPS states:
\setbox0=\hbox{a non-normalizable}
\begin{equation}\label{FromPSToVectorInH}
\fbox{\vbox{\hbox{pure spinors}\hbox{$\lambda_3,\lambda_1$}\hbox{\vbox to \ht0 {}}}} 
\vbox{\hbox{\LARGE $\mapsto$}\hbox{\vbox to \ht0 {}}}
\fbox{\vbox{\hbox{a non-normalizable}
            \hbox{vector in ${\cal H}$ which}
            \hbox{we call $\langle \mbox{plug}|$}}}
\end{equation}
It would be interesting to describe this map explicitly. 
The non-normalizable vector in ${\cal H}$ on the right
hand side of (\ref{FromPSToVectorInH}) is obviously not invariant
under $psu(2,2|4)$ (it belongs to an irreducible representation). But
it transforms covariantly under $so(1,4)\oplus so(5)\subset psu(2,2|4)$,
in the sense that the action of $so(1,4)\oplus so(5)$ on the right
hand side of (\ref{FromPSToVectorInH}) agrees with the action of 
$so(1,4)\oplus so(5)$
on the left hand side of (\ref{FromPSToVectorInH}). 

Another way of thinking about $\langle \mbox{plug}|$ is in terms of
the cohomology of the operator:
\begin{equation}\label{LambdaTInConclusions}
{1\over z}\lambda^{\alpha}_3 t^3_{\alpha} + z\lambda^{\dot{\alpha}}_1 t^1_{\dot{\alpha}}
\end{equation}
acting on the BPS representation ${\cal H}$ (more precisely, the ${\bf g}_{\bar{0}}$-invariant
tensor product of ${\cal H}'$ with the space of polynomials
of $\lambda_3,\lambda_1$).  Our results imply that
the second cohomology of this operator is nontrivial\footnote{
A similar (but different) cohomology problem 
was considered in 
\cite{Kinney:2005ej,Bhattacharya:2008zy,Bhattacharya:2008bja}.}, represented
by the cocycle (\ref{FromPSToVectorInH}). 

Notice that this provides a purely representation-theoretic
characterization of the linearized SUGRA spectrum on $AdS_5\times S^5$.
Indeed, the question of the existence of the excitation transforming
in the representation ${\cal H}$ is reduced to the calculation of the
cohomology of the operator (\ref{LambdaTInConclusions}), which is defined
in terms of the generators $t_a$ of the representation ${\cal H}$.

There is also another example of a plug, a plug of the ghost number 1.
Consider the Wilson line in the adjoint representation. The cohomology
of (\ref{LambdaTInConclusions}) in the adjoint representation is 
nontrivial and is represented by:
\begin{equation}\label{AdjointRepresentative}
{1\over z}\lambda^{\alpha}_3 t^3_{\alpha} - z\lambda^{\dot{\alpha}}_1 t^1_{\dot{\alpha}}
\end{equation}
This is obviously a $\lambda$-dependent vector in the adjoint representation, 
of the ghost number $1$. One can verify that this is annihilated by 
(\ref{LambdaTInConclusions}); note the relative minus sign of the second
term in (\ref{AdjointRepresentative}). Therefore we can take 
(\ref{AdjointRepresentative}) as a plug, and consider:
\[
V(\tau^+,\tau^-|z)
=
\mbox{Str}\left(\left( 
{1\over z}\lambda^{\alpha}_3 t^3_{\alpha} - z\lambda^{\dot{\alpha}}_1 t^1_{\dot{\alpha}}
\right)
P\exp\left( -\int_{\infty}^{(\tau^+,\tau^-)} J[z] \right) 
\Psi^{(\infty)} \right)
\]
At $z=1$ the corresponding integrated vertex operator is the density of the
local conserved charge $\mbox{Str}((j_+d\tau^+ - j_-d\tau^-) \Psi^{(\infty)})$.
We will prove in Appendix \ref{sec:GhostNumberOne} that (\ref{AdjointRepresentative})
is the only example of the endpoint cohomology at ghost number 1. 
In particular, there is no nontrivial cohomology for representations other
than the adjoint.

With these notations the 2-point vertex (\ref{TwoPointVertex}) reads:
\begin{eqnarray}
&&V^{2-pt}((\tau_2^+,\tau_2^-),(\tau_1^+,\tau_1^-)|z) =
\\ 
&&=
\left\langle \mbox{plug}(\tau_2^+,\tau_2^-) \left| 
P\exp\left( -\int_{(\tau_1^+,\tau_1^-)}^{(\tau_2^+,\tau_2^-)} J[z] \right) 
\right| \mbox{plug}(\tau_1^+,\tau_1^-) \right\rangle
\nonumber
\end{eqnarray}
(But as we discussed at the end of Section \ref{sec:TwoPoint} this 
must be BRST exact.)

\section*{Acknowledgments}
I want to thank Y.~Aisaka, N.J.~Berkovits, V.~Serganova and B.C.~Vallilo 
for many useful discussions.
This research was supported by the Sherman Fairchild 
Fellowship and in part
by the RFBR Grant No.  06-02-17383 and in part by the 
Russian Grant for the support of the scientific schools
NSh-8065.2006.2.
Part of this work was done during the Workshop 
``Non-perturbative methods in strongly-coupled gauge theories'',
at the  Galileo Galilei Institute for Theoretical Physics in Florence.
I would like to thank the organizers of this workshop for their hospitality.
Another part of this work was done during the Monsoon Workshop in TIFR, Mumbai.
I want to thank the organizers and the staff members at TIFR for their
hospitality. Another part of this work was done during my stay at the
IFT S\~ao Paulo; I want to thank N.J.~Berkovits for his hospitality.
Another part of this work was done during the workshop 
``Fundamental Aspects of Superstring Theory''
at KITP Santa Barbara; 
I want to thank the organizers of the workshop for their hospitality.

\appendix

\section{Cohomology at ghost number one}
\label{sec:GhostNumberOne}
In this  section we will prove that the only  cohomology at ghost
number 1 are the global  $psu(2,2|4)$ conserved charges.

\subsection{Global conserved charges and BRST cohomology}
The conserved charges are the descendants of the cohomology classes
of the ghost number 1. The ``standard'' local conserved charges correspond
to the global symmetries $PSU(2,2|4)$. They descend from the following operator:
\begin{equation}
\mbox{Ad}(g). (\lambda_3^{\alpha} t^3_{\alpha} - \lambda_1^{\dot{\alpha}} t^1_{\dot{\alpha}})
\end{equation}
In other words, we have the following cohomology class of the ghost number
one in the adjoint representation of ${\bf g}$:
\begin{equation}\label{AdjointClass}
\lambda_3^{\alpha} t^3_{\alpha} - \lambda_1^{\dot{\alpha}} t^1_{\dot{\alpha}}
\end{equation}
In this section we will prove that there are no nontrivial cohomology
classes of the ghost number 1 in the {\em covariant complex}, 
in representations other than the adjoint.

\subsection{Cohomology classes at ghost number 1: the defining equations.}
Fix a representation ${\cal F}$ of ${\bf g} = psu(2,2|4)$.
We will assume two things about ${\cal F}$:
\begin{itemize}

\item   as a representation of ${\bf g}$ it is irreducible

\item   as a representation of ${\bf g}_{even}$ it is completely reducible,
        {\it i.e.} decomposes into the direct sum of irreducible representations

\end{itemize}
We will write a representative of the cohomology class in the following way:
\begin{equation}
\lambda_3^{\alpha}V_{\alpha} + \lambda_1^{\dot{\alpha}}\tilde{V}_{\dot{\alpha}}
\end{equation}
The condition of ${\bf g}_0$-invariance says
that $V_{\alpha}$ and $\tilde{V}_{\dot{\alpha}}$ should define intertwining
operators of ${\bf g}_0$:
\begin{eqnarray}
V & \in & \mbox{Hom}_{{\bf g}_0}({\bf g}_3, {\cal F}), \\
\tilde{V} & \in & \mbox{Hom}_{{\bf g}_0}({\bf g}_1, {\cal F})
\end{eqnarray}
In other words:
\begin{eqnarray}
&& t^0_{[\rho\sigma]} V_{\alpha} = {f_{[\rho\sigma]\alpha}}^{\beta}V_{\beta}
\label{FirstCovarianceCondition}\\
&& t^0_{[\rho\sigma]} V_{\dot{\alpha}} = {f_{[\rho\sigma]\dot{\alpha}}}^{\dot{\beta}}
V_{\dot{\beta}}
\label{SecondCovarianceCondition}
\end{eqnarray}
--- the conditions of ${\bf g}_0$-covariance. 
The conditions for being annihilated by $Q$ are:
\begin{eqnarray}\label{ConditionThatQAnnihilates}
&& t^3_{\alpha}V_{\beta}+t^3_{\beta}V_{\alpha} = {f_{\alpha\beta}}^{\mu} A_{\mu} 
\label{FirstIdentity}\\
&& t^1_{\dot{\alpha}} \tilde{V}_{\dot{\beta}} + t^1_{\dot{\beta}}\tilde{V}_{\dot{\alpha}} =
{f_{\dot{\alpha}\dot{\beta}}}^{\mu}\tilde{A}_{\mu} 
\label{SecondIdentity}\\
&& t^3_{\alpha}\tilde{V}_{\dot{\beta}} + t^1_{\dot{\beta}}V_{\alpha} = 0
\label{ThirdIdentity}
\end{eqnarray}
For example, the class (\ref{AdjointClass}) is represented by:
\begin{eqnarray}
V_{\alpha} & = & t^3_{\alpha}
\nonumber
\\
\tilde{V}_{\dot{\alpha}} & = & - t^1_{\dot{\alpha}} 
\label{OnlySolution}\\
A_{\mu} =  -\tilde{A}_{\mu} & = & 2t^2_{\mu}
\nonumber
\end{eqnarray}
We consider the solutions of (\ref{ConditionThatQAnnihilates}) trivial if they are of the form:
\begin{equation}
V_{\alpha} = t^3_{\alpha}\Phi\;,\;\;
V_{\dot{\alpha}} = t^1_{\dot{\alpha}}\Phi
\end{equation}
where $t_0^{[\mu\nu]}\Phi=0$.

We want to prove the following:

\vspace{10pt}

\noindent {\em Theorem:} Nontrivial solutions to Eqs. (\ref{FirstIdentity} -- \ref{ThirdIdentity})
exist only when ${\cal F}$ is the adjoint representation of ${\bf g}$, and 
are given by (\ref{OnlySolution}) up to adding a trivial solution. There are no other 
nontrivial solutions.

\vspace{10pt}

\noindent
We will now proceed to prove this.

\subsection{Cohomology classes at ghost number 1: consequences of the defining equations}
Acting on  (\ref{FirstIdentity}) by $t^1_{\dot{\beta}}$ we get:
\begin{equation}
\left({f_{\dot{\beta}\alpha}}^{[\rho\sigma]} t^0_{[\rho\sigma]} V_{\beta} -
t^3_{\alpha}t^1_{\dot{\beta}}V_{\beta} \right) + (\alpha\leftrightarrow\beta) 
= {f_{\alpha\beta}}^{\mu} t^1_{\dot{\beta}} A_{\mu}
\end{equation}
This with Eqs. (\ref{FirstCovarianceCondition}) and (\ref{ThirdIdentity}) implies:
\begin{equation}
\left({f_{\dot{\beta}\alpha}}^{[\rho\sigma]} {f_{[\rho\sigma]\beta}}^{\gamma} V_{\gamma}
+ (\alpha\leftrightarrow\beta)\right) + {f_{\alpha\beta}}^{\mu}t^2_{\mu} 
\tilde{V}_{\dot{\beta}} 
= {f_{\alpha\beta}}^{\mu}t^1_{\dot{\beta}} A_{\mu}
\end{equation}
This with the Jacobi identity for $ff$ implies:
\begin{equation}\label{fV}
 {f_{\mu\dot{\beta}}}^{\alpha}V_{\alpha} = 
t^2_{\mu}\tilde{V}_{\dot{\beta}} - t^1_{\dot{\beta}}A_{\mu} 
\end{equation}
Similarly we have:
\begin{equation}\label{ftV}
 {f_{\mu\beta}}^{\dot{\alpha}}\tilde{V}_{\dot{\alpha}} = 
t^2_{\mu}V_{\beta} - t^3_{\beta}\tilde{A}_{\mu} 
\end{equation}
Let us act on (\ref{fV}) by ${f^{\dot{\gamma}\dot{\beta}}}_{\nu}t^1_{\dot{\gamma}}$:
\begin{eqnarray}
{f^{\dot{\gamma}\dot{\beta}}}_{\nu}
{f_{\mu\dot{\beta}}}^{\alpha} t^1_{\dot{\gamma}} V_{\alpha}  & = &
{f^{\dot{\gamma}\dot{\beta}}}_{\nu} t^1_{\dot{\gamma}} t^2_{\mu} \tilde{V}_{\dot{\beta}}
- {f^{\dot{\gamma}\dot{\beta}}}_{\nu} t^1_{\dot{\gamma}} t^1_{\dot{\beta}} A_{\mu}
= \\
& = & 
{f^{\dot{\gamma}\dot{\beta}}}_{\nu} {f_{\dot{\gamma}\mu}}^{\alpha} 
t^3_{\alpha} \tilde{V}_{\dot{\beta}} 
+ 
{1\over 2} {f^{\dot{\gamma}\dot{\beta}}}_{\nu} {f_{\dot{\gamma}\dot{\beta}}}^{\lambda}
t^2_{\mu} \tilde{A}_{\lambda}
-
{1\over 2} {f^{\dot{\gamma}\dot{\beta}}}_{\nu} {f_{\dot{\gamma}\dot{\beta}}}^{\lambda}
t^2_{\lambda} A_{\mu}
\nonumber 
\end{eqnarray}
This and (\ref{ThirdIdentity}) implies:
\begin{equation}\label{FirstRelationOfAs}
t^2_{\mu} \tilde{A}_{\nu} - t^2_{\nu} A_{\mu} = 0
\end{equation}
Let us denote: $B_{\mu} = A_{\mu} + \tilde{A}_{\mu}$. We have:
\begin{equation}
t^2_{[\mu} B_{\nu]} = 0
\end{equation}
Note that the gauge transformation 
\begin{equation} 
\delta V_{\alpha} = t^3_{\alpha} \Phi\;\;, \;\;\;
\delta\tilde{V}_{\dot{\alpha}} = t^1_{\dot{\alpha}} \Phi
\end{equation}
where $\Phi$ is ${\bf g}_{\bar{0}}$-invariant leads to:
\begin{equation}
\delta B_{\mu} = t^2_{\mu}\Phi
\end{equation}
Therefore we should think of $B_{\mu}$ as an element of $H^1({\bf g}_{even}, {\bf g}_{\bar{0}}, {\cal F})$. 
But this cohomology group is zero because $H^1({\bf g}_{even},{\cal F})=0$ (notice that the
Serre-Hochschild  spectral sequence for ${\bf g}_{even}\subset {\bf g}$ has $E_2^{p,0} = H^p({\bf g},{\bf g}_{even}, {\cal F})$
and $d_2$ acts from $E_2^{p,q}$ to $E_2^{p+2,q-1}$).
Therefore we should be able to gauge away $B_{\mu}$. Let us therefore assume:
\begin{equation}
A_{\mu} = -\tilde{A}_{\mu}
\end{equation}
Note that this equation and (\ref{FirstRelationOfAs}) implies:
\begin{equation}\label{AntiSymmetryOfF}
t^2_{\mu}A_{\nu}+t^2_{\nu}A_{\mu}=0
\end{equation}
Now we can rewrite (\ref{fV}) and (\ref{ftV}) as follows:
\begin{eqnarray}
&& {f_{\mu\dot{\beta}}}^{\alpha} V_{\alpha} = t^2_{\mu} \tilde{V}_{\dot{\beta}} - 
t^1_{\dot{\beta}} A_{\mu}  
\label{FirstFV}
\\
&& {f_{\mu\beta}}^{\dot{\alpha}} \tilde{V}_{\dot{\alpha}} =
t^2_{\mu}V_{\beta} + t^3_{\beta} A_{\mu}
\label{SecondFV}
\end{eqnarray}
Let us define $F_{\mu\nu}$ by the following equation:
\begin{equation}
F_{\rho\sigma} = t^2_{\rho}A_{\sigma}
\end{equation}
Eq. (\ref{AntiSymmetryOfF}) implies that $F_{\rho\sigma}$ is antisymmetric: $F_{\rho\sigma} = - F_{\sigma\rho}$. We get:
\begin{eqnarray}
t^2_{\lambda}F_{\mu\nu} & = & t^2_{\lambda} t^2_{\mu} A_{\nu} =
{f_{\lambda\mu}}^{[\rho\sigma]} t^0_{[\rho\sigma]} A_{\nu} +
t^2_{\mu} t^2_{\lambda} A_{\nu} = 
\nonumber 
\\
& = & {f_{\lambda [\mu |}}^{[\rho\sigma]} 
{f_{[\rho\sigma]|\nu]}}^{\kappa} A_{\kappa} -
t^2_{[\mu} t^2_{\nu]} A_{\lambda} =
\nonumber
\\
& = & -{1\over 2} {f_{\mu\nu}}^{[\rho\sigma]} {f_{[\rho\sigma]\lambda}}^{\kappa} A_{\kappa} -
{1\over 2} {f_{\mu\nu}}^{[\rho\sigma]} {f_{[\rho\sigma]\lambda}}^{\kappa} A_{\kappa} =
\nonumber
\\
& = & - {f_{\mu\nu}}^{[\rho\sigma]} {f_{[\rho\sigma]\lambda}}^{\kappa} A_{\kappa}
\label{T2OnF}
\end{eqnarray}
Therefore $F_{\mu\nu}$ can be expressed in terms of 
$G_{[\mu\nu]}$ and $M_{\mu\nu}$ which are defined by this equation:
\begin{equation}
F_{\mu\nu} = {f_{\mu\nu}}^{[\rho\sigma]} G_{[\rho\sigma]} + M_{\mu\nu}
\end{equation}
where $M_{\mu\nu} = -M_{\nu\mu}$ is nonzero only when $\mu$ is tangent to $AdS_5$ and $\nu$ is 
tangent to $S_5$, or vice versa, and:
\begin{align}
t^2_{\lambda} M_{\mu\nu} =&\; 0
\\   
t^2_{\lambda} G_{[\mu\nu]} =&\; {f_{\lambda[\mu\nu]}}^{\kappa} A_{\kappa}
\end{align}
Then the covariance under ${\bf g}_{\bar{0}}$ implies that $M_{\mu\nu} = 0$.
This means that the linear space formed by $A_{\mu}$ and $G_{[\mu\nu]}$ is closed under
the action of ${\bf g}_{even}$, and is in fact the adjoint representation of ${\bf g}_{even}$
(where $A_{\mu}$ corresponds to $2t_{\mu}$ 
and $G_{[\mu\nu]}$ corresponds to $2t_{[\mu\nu]}$).
This is already close to what we wanted to prove. But we have to also tame the
expressions of this form:
\begin{equation}
t^3_{\alpha} t^1_{\dot{\beta}} t^1_{\dot{\gamma}} t^3_{\delta} A_{\mu}
\;\; , \;\;\;
t^1_{\dot{\alpha}} t^3_{\beta} t^3_{\gamma} G_{[\mu\nu]}
\;\; , \;\;\; etc.
\end{equation}
For this purpose, let us use (\ref{FirstFV}) and (\ref{SecondFV}) in this expression:
\begin{eqnarray}
&& t^3_{\alpha}t^1_{\dot{\beta}} A_{\mu} - t^1_{\dot{\beta}}t^3_{\alpha} 
A_{\mu} =
\nonumber
\\
&& = t^3_{\alpha}(t^2_{\mu} \tilde{V}_{\dot{\beta}} - 
{f_{\mu\dot{\beta}}}^{\gamma} V_{\gamma} )+
t^1_{\dot{\beta}}  (t^2_{\mu} V_{\alpha} - 
{f_{\mu\alpha}}^{\dot{\gamma}}
\tilde{V}_{\dot{\gamma}} ) =
\nonumber
\\
&& = - {f_{\mu\alpha}}^{\dot{\gamma}} 
(t^1_{\dot{\gamma}} \tilde{V}_{\dot{\beta}} 
+ t^1_{\dot{\beta}}\tilde{V}_{\dot{\gamma}}) 
- {f_{\mu\dot{\beta}}}^{\gamma} 
(t^3_{\alpha}V_{\gamma} + t^3_{\gamma}V_{\alpha}) =
\nonumber
\\
&& = {f_{\dot{\beta}\mu}}^{\gamma}{f_{\alpha\gamma}}^{\nu} A_{\nu} 
   - {f_{\alpha\mu}}^{\dot{\gamma}}{f_{\dot{\gamma}\dot{\beta}}}^{\nu} A_{\nu}
\end{eqnarray}
On the other hand, the combination 
$t^3_{\alpha}t^1_{\dot{\beta}} A_{\mu} + t^1_{\dot{\beta}}t^3_{\alpha} A_{\mu}$
can be calculated using the ${\bf g}_0$-invariance. This implies:
\begin{equation}
t^3_{\alpha} t^1_{\dot{\beta}} A_{\mu} =  
{f_{\dot{\beta}\mu}}^{\gamma} {f_{\alpha\gamma}}^{\nu} A_{\nu}
\label{ReduceA}
\end{equation}
We will also use this:
\begin{eqnarray}
&&t^1_{\dot{\alpha}}F_{\mu\nu} = t^1_{\dot{\alpha}} t^2_{\mu} A_{\nu} =
{1\over c} {f_{\mu}}^{\gamma\delta} t^1_{\dot{\alpha}}
t^3_{\gamma} t^3_{\delta} A_{\nu} =
\nonumber
\\
&& = {1\over c} {f_{\mu}}^{\gamma\delta} 
{f_{\dot{\alpha}\gamma}}^{[\rho\sigma]} t^0_{[\rho\sigma]} 
t_{\delta}^3 A_{\nu} - {1\over c} {f_{\mu}}^{\gamma\delta}
t^3_{\gamma} t^1_{\dot{\alpha}} t^3_{\delta} A_{\nu} =
\nonumber
\\
&& = {1\over c}(fff)(t^3 A)
\label{ThreeStructureConstants}
\end{eqnarray}
Here we used the schematic notation $(fff)$ for a product of  three
structure constants with some indices contracted, and $c$ is
determined from 
${f_{\mu}}^{\alpha\beta}{f_{\alpha\beta}}^{\nu} = c\delta_{\mu}^{\nu}$.
The subspace of ${\cal F}$ generated by acting on $A$ and $G$ by finitely many
$t^3$ and $t^1$ is finite-dimensional. Indeed, using 
(\ref{ReduceA}) and (\ref{ThreeStructureConstants})
we can prove that it is generated as a linear space by expressions of the form:
\begin{eqnarray}
&& t^1_{[\dot{\alpha}_1} \cdots t^1_{\dot{\alpha}_k]} A_{\mu}\;,\;\;
t^3_{[\alpha_1} \cdots t^3_{\alpha_k]} A_{\mu}
\;\;\;
(k\geq 0)
\label{SubspaceGeneratorsA}
\\
& \mbox{and} & G_{[\mu\nu]}
\label{SubspaceGeneratorsG}
\end{eqnarray}
where the square brackets stand for the antisymmetrization of the indices
(for example $t^3_{[\alpha} t^3_{\beta]} A_{\mu}$ stands
for $(t^3_{\alpha}t^3_{\beta} - t^3_{\beta} t^3_{\alpha})A_{\mu}$). 
Eqs. (\ref{ReduceA}) and (\ref{ThreeStructureConstants}) 
imply that this subspace is closed 
under the action of  ${\bf g}$. 
Because ${\cal F}$ is assumed to be irreducible, we conclude that
${\cal F}$ is generated by 
(\ref{SubspaceGeneratorsA}),(\ref{SubspaceGeneratorsG}).
Because of
the antisymmetrization of the indices of $t^3$ and $t^1$ 
there are only finitely many linearly independent
expressions of the form (\ref{SubspaceGeneratorsA}).
This proves that ${\cal F}$ is a finite-dimensional space. 

\vspace{10pt}

\noindent
The subspace in ${\cal F}$ generated by:
\begin{equation}\label{SmallSetOfGenerators}
A_{\mu}\;,\;\;
G_{[\mu\nu]}\;,\;\;
t_{\alpha}A_{\mu}\;,\;\;
t_{\dot{\alpha}}A_{\mu}
\end{equation}
is closed under ${\bf g}_{even}$.
Let us denote this space ${\cal L}$. 
Obviously ${\cal L}\subset {\cal F}$.

\vspace{10pt}

\noindent
{\em Theorem.}
${\cal L} = {\cal F}$. 

\vspace{5pt}

\noindent
{\em Proof.} One can see that for any element $v$ of $\cal H$ 
({\it i.e.} a finite linear combination of expressions of the form 
(\ref{SubspaceGeneratorsA})) there is a number $p$ such that for any $q>p$
and any $q$ elements $\xi_1,\ldots,\xi_q$ of ${\bf g}_{even}$ we get:
\begin{equation}\label{EventuallyReachL}
\xi_1\cdots \xi_q \; v \in {\cal L}
\end{equation}
Indeed, let us consider for example acting by $\xi\in {\bf g}_{\bar{2}}$ on
expressions of the form $t^3_{[\alpha_1} \cdots t^3_{\alpha_k]} A_{\mu}$.
Let us define the degree of such an expression by the following formula:
\begin{equation}
\mbox{deg } t^3_{[\alpha_1} \cdots t^3_{\alpha_k]} A_{\mu} = 
\mbox{deg } t^1_{[\dot{\alpha}_1} \cdots t^1_{\dot{\alpha}_k]} A_{\mu} = k
\end{equation}
More precisely, we introduce a filtration of $\cal H$ saying that
$F^k {\cal H}$ consists of all the elements of ${\cal H}$ which can be
written as linear combinations of the expressions of the
form (\ref{SubspaceGeneratorsA}) of the degree less or equal $k$.
Using Eqs. (\ref{ReduceA}) and (\ref{ThreeStructureConstants}) 
we derive that for $k>1$:
\begin{equation}
t^2_{\mu} F^k {\cal H}  \subset F^{k-1} {\cal H}
\end{equation}
This implies (\ref{EventuallyReachL}).
Because of the assumption that ${\cal F}$ is completely reducible as a
representation of ${\bf g}_{even}$, Eq. (\ref{EventuallyReachL}) implies
that ${\cal L} = {\cal F}$.

\vspace{10pt}

\noindent We conclude that ${\cal F}$ is in fact generated by 
expressions (\ref{SmallSetOfGenerators}). Note that 
the linear space generated by (\ref{SmallSetOfGenerators}) consists
of the even subspace generated by $A_{\mu}$ and $G_{[\mu\nu]}$, and odd subspace
generated by $t_{\alpha}A_{\mu}, t_{\dot{\alpha}}A_{\mu}$. 
The even subspace is the same as in the 
adjoint representation. Therefore the odd space should be also the same. 

This proves that $\cal H$ is the adjoint representation of ${\bf g}=psu(2,2|4)$.

\subsection{Relaxing the requirement that ${\cal F}$ is irreducible}
We have argued that the subspace generated by (\ref{SubspaceGeneratorsA}) in fact coincides with 
${\cal F}$, based on ${\cal F}$ being an irreducible representation of ${\bf g}$. This requirement 
can be replaced with the requirement that $H^1({\bf g},{\cal F}) = 0$. Suppose that (\ref{SubspaceGeneratorsA})
generate a smaller subspace ${\cal F}_{A+G}\subset {\cal F}$. 
Let us denote $v$ and $\tilde{v}$ the projections of $V$ and $\tilde{V}$ on ${\cal F}/{\cal F}_{A+G}$:
\begin{equation}
v=V \;\; \mbox{mod} \;\; {\cal F}_{A+G}
\end{equation}
Then (\ref{SecondIdentity}) implies:
\begin{equation}
t^3_{\alpha}v_{\beta}+t^3_{\beta}v_{\alpha} \; = \;
t^1_{\dot{\alpha}} \tilde{v}_{\dot{\beta}} + t^1_{\dot{\beta}}\tilde{v}_{\dot{\alpha}} \; = \;
t^3_{\alpha}\tilde{v}_{\dot{\beta}} + t^1_{\dot{\beta}}v_{\alpha} \; = \; 0
\end{equation}
These equations imply that $(v_{\alpha},\tilde{v}_{\dot{\beta}})$ form the spinor representation of ${\bf g}_{even}$:
\begin{eqnarray}
t_m v_{\alpha} & = & {f_{m\alpha}}^{\dot{\beta}} \tilde{v}_{\dot{\beta}}
\label{GEvenInvariance1}
\\
t_m \tilde{v}_{\dot{\alpha}} & = & {f_{m\dot{\alpha}}}^{\beta} v_{\beta}
\label{GEvenInvariance2}
\end{eqnarray}
We will now explain, using (\ref{GEvenInvariance1}) and (\ref{GEvenInvariance2}), that $v$ can be gauged away if  
$H^1({\bf g},{\bf g}_{even},{\cal F})=0$. We will also explain that $H^1({\bf g},{\bf g}_{even},{\cal F})=0$ if 
$H^1({\bf g},{\cal F}) = 0$.

More generally, let us consider the relative Lie algebra cohomology complex 
$C^{\bullet}({\bf g}, {\bf g}_{even}, {\cal F})$. The cochains are tensors with spinor indices satisfying:
\begin{equation}
t^2_m v_{\alpha_1\ldots\alpha_p\; \dot{\beta}_1\ldots\dot{\beta}_q} =
p {f_{m(\alpha_1}}^{\dot{\alpha}_1}
v_{\alpha_2\ldots\alpha_p)\; \dot{\alpha}_1\dot{\beta}_1\ldots\dot{\beta}_q} 
+
q {f_{m(\dot{\beta}_1|}}^{{\beta}_1}
v_{\beta_1\alpha_1\ldots\alpha_p|\;\dot{\beta}_2\ldots\dot{\beta}_q)} 
\end{equation}
The differential in relative cohomology is: 
\begin{equation}
(Qv)_{\alpha_1\cdots\alpha_p \dot{\beta}_1\cdots \dot{\beta}_q} = 
t^3_{(\alpha_1}v_{\alpha_2\cdots\alpha_p) \dot{\beta}_1\cdots \dot{\beta}_q} + 
t^1_{(\dot{\beta}_1|}v_{\alpha_1\cdots\alpha_p | \dot{\beta}_2\cdots\dot{\beta}_q)}
\end{equation}
This can be thought of as a distant relative of the pure spinor BRST complex.
The difference is that a stronger covariance condition is imposed 
(${\bf g}_{even}\supset {\bf g}_{\bar{0}}$) and also no constraints on the ghost variables. 
We will now explain that the BRST cohomology of the ${\bf g}_{even}$-covariant complex 
is  zero for large enought quantum numbers, unlike the cohomology of 
the ``normal'' BRST complex (which is only ${\bf g}_{\bar{0}}$-covariant).

Indeed, this relative cohomology 
is related to $H^{\bullet}({\bf g},{\cal F})$ by the Serre-Hochschild spectral sequence. Namely 
\begin{equation}
H^p({\bf g},{\bf g}_{even},{\cal F}) = E_2^{p,0}
\end{equation}
The differential $d_r$ acts from $E_r^{p,q}$ to $E_r^{p+r, q+1-r}$.
In particular, $E_{r}^{1,0}$ is related to $E_{r}^{1+r,1-r}$.
\begin{equation}
E_r^{1-r, -1+r} \stackrel{d_r}{\longrightarrow} E_r^{1,0}
\stackrel{d_r}{\longrightarrow} E_r^{1+r, 1-r}
\end{equation}
This means that $E_2^{1,0}$ cannot cancel with anything and therefore $H^1({\bf g},{\cal F}) = 0$ 
implies that $H^1({\bf g},{\bf g}_{even},{\cal F}) = 0$. 

Similarly, vanishing of $H^2({\bf g},{\cal F})$ and $H^1({\bf g}_{even},{\cal F})$ implies vanishing of 
$H^2({\bf g},{\bf g}_{even},{\cal F})$. Indeed, the action of $d_2$ is:
\begin{equation}
E_2^{0, 1} \stackrel{d_2}{\longrightarrow} E_2^{2,0}
\stackrel{d_2}{\longrightarrow} (0 = E_2^{4, -1})
\end{equation}
where:
\begin{align}
E_2^{2,0} = &\; H^2({\bf g}, {\bf g}_{even}, {\cal F})
\\    
E_1^{0,1} = &\; H^1({\bf g}_{even}, {\cal F})
\end{align}
Therefore:
\begin{align}  
F^2H^2({\bf g},{\cal F}) = &\; E_2^{2,0}/\mbox{Im}\; (d_2:E_2^{0,1} \to E_2^{2,0})
\end{align}

\section{Shapiro's lemma}\label{sec:Shapiro}
This section is a brief review of the Shapiro's lemma applied to Lie 
superalgebras:
\begin{equation}
H^n({\bf g},\; \mbox{Hom}_{\bf C}({\cal H}, \mbox{Hom}_{{\cal U}{\bf h}}({\cal U}{\bf g},A)))
= 
H^n({\bf h},\; \mbox{Hom}_{\bf C}({\cal H}|_{\bf h}, A))
\end{equation}
We will follow \cite{Knapp}. 

\subsection{Relation between cohomology and Ext}
The proof starts with pointing out
the relation between the cohomology and the Ext group:
\begin{equation}\label{FromCohomologyToExt}
H^n\left({\bf g},\mbox{Hom}_{\bf C}(M,N)\right) 
= \mbox{Ext}^n_{{\cal U}{\bf g}}(M,N)
\end{equation}
This is proven as follows. By definition $\mbox{Ext}^n_{{\cal U}{\bf g}}(M,N)$ is computed using 
the projective resolution $\ldots \rightarrow P_M^1 \rightarrow P_M^0 \rightarrow M\rightarrow 0$ of $M$, as a module
over ${\cal U}{\bf g}$. Given such a projective resolution,  $\mbox{Ext}^n_{{\cal U}{\bf g}}(M,N)$ is identified
with the $n$-th cohomology group of the complex:
\begin{equation}\label{HomPMN}
\ldots
\mbox{Hom}_{{\cal U}{\bf g}} (P_M^{n-1}, N) \rightarrow
\mbox{Hom}_{{\cal U}{\bf g}} (P_M^{n}, N)  \rightarrow
\mbox{Hom}_{{\cal U}{\bf g}} (P_M^{n+1}, N)  \rightarrow 
\ldots
\end{equation}
This is the complex of {\em vector spaces}, the spaces of invariants in the 
modules $\mbox{Hom}_{\bf C}(P^n_M,N)$. But in fact $\mbox{Hom}_{\bf C}(P^n_M,N)$ are  injective 
${\cal U}{\bf g}$-modules. Therefore the following complex is an injective resolution 
of $\mbox{Hom}_{\bf C}(M,N)$:
\begin{equation}\label{HomCPMN}
\ldots
\mbox{Hom}_{{\bf C}} (P_M^{n-1}, N) \rightarrow
\mbox{Hom}_{{\bf C}} (P_M^{n}, N)  \rightarrow
\mbox{Hom}_{{\bf C}} (P_M^{n+1}, N)  \rightarrow 
\ldots
\end{equation}
Therefore the cohomologies of (\ref{HomPMN}) are identified with the Lie algebra
cohomologies $H^n\left({\bf g},\mbox{Hom}_{\bf C}(M,N)\right)$. It remains to prove that $\mbox{Hom}_{\bf C}(P^n_M,N)$ 
are  injective ${\cal U}{\bf g}$-modules. It turns out that if $P$ is projective, then
$\mbox{Hom}_{\bf C}(P,N)$ is injective. This is equivalent to the statement that the
following contravariant functor:
\begin{equation}\label{Functor}
W\mapsto \mbox{Hom}_{{\cal U}{\bf g}} 
\left(W, \mbox{Hom}_{\bf C}(P,N)\right)
\end{equation}
is exact. Notice that there is a canonical isomorphism:
\begin{align}
\mbox{Hom}_{{\cal U}{\bf g}} 
\left(W, \mbox{Hom}_{\bf C}(P,N)\right) 
\simeq & \;
\mbox{Hom}_{{\cal U}{\bf g}} 
\left(P, \mbox{Hom}_{\bf C}(W,N)\right) 
\\   
(f:\; W\to \mbox{Hom}_{\bf C}(P,N)) 
\mapsto & \;
(g:\; P\to \mbox{Hom}_{\bf C}(W,N))
\end{align}
where: 
\begin{equation}
g(p)(w) = (-)^{\bar{p}\bar{w}} f(w)(p)
\end{equation}
We have to verify that so defined $g$ indeed belongs to 
$\mbox{Hom}_{{\cal U}{\bf g}} 
\left(P, \mbox{Hom}_{\bf C}(W,N)\right)$; what has to be verified is the invariance of $g$
under ${\bf g}$. For $\xi\in {\bf g}$, taking into account that used that $\bar{f} = \bar{g}$, we get:
\begin{align}
(\xi . g) (p) (w) = 
&\; \rho_N(\xi) (g (p) (w)) 
- (-)^{\bar{\xi}\;\overline{g(p)}} g (p) (\rho_W(\xi)w) -
\nonumber \\    %
& 
- (-)^{\bar{\xi}\;\bar{g}}g(\rho_P(\xi)p)(w) =
\nonumber \\    %
=&\;(-)^{\bar{p}\bar{w}}
\rho_N(\xi) (f (w) (p)) 
- (-)^{\bar{\xi}\;\overline{g(p)} + \bar{p}(\bar{\xi} + \bar{w})} 
f (\rho_W(\xi)w) (p)  -
\nonumber \\    %
& - (-)^{\bar{\xi}\;\bar{g} + \bar{w}(\bar{p} + \bar{\xi})}
f(w)(\rho_P(\xi)p) = 
\\    %
= &\; (-)^{\bar{p}\bar{w}}\left(
\rho_N(\xi) (f (w) (p))
- (-)^{\bar{\xi}\;\overline{f}} 
f (\rho_W(\xi)w) (p)  - \right)
\nonumber \\    %
& \left.
\quad\quad\quad - (-)^{\bar{\xi}\;\overline{f(w)}}
f(w)(\rho_P(\xi)p)
\right)
\nonumber
\end{align}
This is zero because  of the covariance condition on $f$. It can be similarly
verified that the map $f\mapsto g$ commutes with the action of ${\cal U}{\bf g}$. Therefore
(\ref{Functor}) is naturally equivalent to:
\begin{equation}\label{FunctorExchanged}
W\mapsto \mbox{Hom}_{{\cal U}{\bf g}} 
\left(P, \mbox{Hom}_{\bf C}(W,N)\right)
\end{equation}
which is a composition of the exact functors $W\mapsto \mbox{Hom}_{\bf C}(W,N)$ and 
$V\mapsto\mbox{Hom}_{{\cal U}{\bf g}} (P, V)$. This means that (\ref{Functor}) is an exact functor, and therefore
$\mbox{Hom}_{\bf C}(P,N)$ is an injective ${\cal U}{\bf g}$-module. This concludes the proof of (\ref{FromCohomologyToExt}).

\subsection{Shapiro's lemma for Ext}
\begin{equation}
\mbox{Ext}^n_{{\cal U}{\bf g}}(M, \mbox{Hom}_{{\cal U}{\bf h}}({\cal U}{\bf g},A))
= 
\mbox{Ext}^n_{{\cal U}{\bf h}}(M|_{\bf h},A)
\end{equation}
This is proved by noticing that:
\begin{equation}
\mbox{Hom}_{{\cal U}{\bf g}}\left(
   P^i_M,\mbox{Hom}_{{\cal U}{\bf h}}({\cal U}{\bf g},A)
\right) 
=
\mbox{Hom}_{{\cal U}{\bf h}}\left(
   \left.P^i_M\right|_{{\cal U}{\bf h}}\;,\;A
\right)
\end{equation}
and that $\left.P^i_M\right|_{{\cal U}{\bf h}}$ is a projective resolution for the restriction of $M$ to ${\cal U}{\bf h}$,
because ${\cal U}{\bf g}$ is projective (in fact free) as an  ${\cal U}{\bf h}$ module\footnote{therefore
the restriction of a projective module from ${\cal U}{\bf g}$ to ${\cal U}{\bf h}$ is a projective module over ${\cal U}{\bf h}$; to see this observe that
projective modules are the same as free summands of free modules}.

\providecommand{\href}[2]{#2}\begingroup\raggedright\endgroup

\end{document}